\documentclass[twocolumn,english,superscriptaddress,prl,amsmath,amssymb,floatfix,longbibliography]{revtex4-1}
\usepackage[T1]{fontenc}
\usepackage{xcolor}
\usepackage{pdfcolmk}
\usepackage{babel}
\usepackage{amsmath}
\usepackage{graphicx}
\usepackage{esint}

\PassOptionsToPackage{normalem}{ulem}
\usepackage{ulem}
\usepackage[unicode=true,pdfusetitle,
 bookmarks=true,bookmarksnumbered=false,bookmarksopen=false,
 breaklinks=false,pdfborder={0 0 0},pdfborderstyle={},backref=false,colorlinks=true]
 {hyperref}

\makeatletter

\providecolor{lyxadded}{rgb}{0,0,1}
\providecolor{lyxdeleted}{rgb}{1,0,0}

\DeclareRobustCommand{\lyxsout}[1]{\ifx\\#1\else\sout{#1}\fi}

\usepackage{dcolumn}
\usepackage{bm}
\usepackage{color}
\usepackage{soul}
\usepackage{babel}
\usepackage{amsfonts}
\usepackage{slashed}
\usepackage{enumerate}

\usepackage{babel}

\makeatother

\begin{document}
\global\long\def\sgn{\mathrm{sgn}}%
\global\long\def\ket#1{\left|#1\right\rangle }%
\global\long\def\bra#1{\left\langle #1\right|}%
\global\long\def\sp#1#2{\langle#1|#2\rangle}%
\global\long\def\abs#1{\left|#1\right|}%
\global\long\def\avg#1{\langle#1\rangle}%
\global\long\def\purity{P}%

\title{Optimized Steering: Quantum State Engineering and Exceptional Points}
\author{Parveen Kumar}
\affiliation{Department of Condensed Matter Physics, Weizmann Institute of Science,
Rehovot 7610001, Israel}
\author{Kyrylo Snizhko}
\affiliation{Institute for Quantum Materials and Technologies, Karlsruhe Institute of Technology, 76021 Karlsruhe, Germany}
\affiliation{Department of Condensed Matter Physics, Weizmann Institute of Science,
	Rehovot 7610001, Israel}
\author{Yuval Gefen}
\affiliation{Department of Condensed Matter Physics, Weizmann Institute of Science,
	Rehovot 7610001, Israel}
\author{Bernd Rosenow}
\affiliation{Institut f\"{u}r Theoretische Physik, Universit\"{a}t Leipzig, Br\"{u}derstrasse
	16, 04103 Leipzig, Germany}
\begin{abstract}
The state of a quantum system may be steered towards a predesignated
target state, employing a sequence of weak \emph{blind} measurements (where the detector's readouts are traced out).
Here we analyze the steering of a two-level system using the interplay
of a system Hamiltonian and weak measurements, and show that \emph{any} pure
or mixed state can be targeted. We show that the optimization of such
a steering protocol is underlain by the presence of Liouvillian
exceptional points. More specifically, for high purity target states,
optimal steering implies purely relaxational dynamics marked by
a second-order exceptional point, while for low purity target states,
it implies an oscillatory approach to the target state. The dynamical phase
transition between these two regimes is characterized by a third-order
exceptional point.
\end{abstract}
\maketitle
Steering of a quantum system towards a pre-designated target state can be achieved either by drive-and-dissipation schemes \citep{Diehl2008a,Kraus2008,Roncaglia2010,Diehl2011,Pechen2011a,Murch2012,Leghtas2013,Shankar2013,Lin2013,Liu2016,Goldman2016,Lu2017,Huang2018,Horn2018}
or through measurement-based protocols \citep{Pechen2006,Roa2006,Roa2007a,Jacobs2010a,Ashhab2010,Roy2019b,Kumar2020}. The former employ a dissipative
environment to relax the quantum system into the target state, while in the latter case relaxation (as well as back-action on the system) is achieved by measurements.
Optimizing the rate of convergence towards the target state is important to render it of practical importance and minimize external perturbations. Refs~\citep{Ticozzi2012,Baumgartner2008} comprise a mathematical analysis of the steady states and
the convergence speed.
Exceptional points (EPs), referring to non-Hermitian degeneracies where two or more eigenvectors of the evolution operator coalesce~\citep{Kato1995,Berry2004,Heiss2012a,Hatano2019a,Minganti2019}, play an important role in a variety of optimization problems~\citep{Lin2016,Lupu2017,Metelmann2018,Partanen2019,Fernandez-Alcazar2020,Wiersig2020c,Chen2021}.
Such degeneracies are of particular interest in dynamics with complex
eigenvalues where unitary dynamics competes with dissipation or gain, and may be generalized from non-Hermitian Hamiltonians to Liouvillian dynamics~\citep{Minganti2020c,Arkhipov2020b}.
In recent years, it has been recognized that operating near an EP
enables unique functionality such as unidirectional invisibility \citep{Lin2011,Regensburger2012,Feng2013,Peng2014}
or enhanced sensitivity \citep{Wiersig2014,Liu2016a,Hodaei2017,Chen2017,Lau2018,Zhang2019b}.

Here, we propose a family of protocols for steering a two-level quantum
system towards  desired target states. The system's initial state
is assumed unknown to us. In our protocol, the quantum system is subject
to both a Hamiltonian evolution and a measurement-induced evolution,
and the combined effect of both  can be described by a Liouvillian
superoperator. The system's target state
is given by the Liouvillian eigenstate having zero eigenvalue, and is
uniquely determined (along with its purity) by the
interplay between the Hamiltonian  and the measurement protocol.
 When optimizing  our steering~\footnote{In this paper, we use the word ``steering'' as the name of a process
that leads an arbitrary initial state of the system to a predesignated
target state. This should not be confused with ``steering'' from
the theory of quantum information and quantum computation which defines
a special kind of nonlocal correlations. } protocol in the sense that the target state is reached as fast as possible, we find that the optimal steering for high purity target states is dominated by the measurement-induced dynamics and described by second-order exceptional points, while optimal protocols  for low
purity target states are dominated by the system-Hamiltonian-induced dynamics.
The transition between these two regimes is characterized by a third-order
EP, where all three nonzero eigenvalues of the Liouvillian superoperator
coalesce, and the optimal convergence dynamics of the system changes
from non-oscillatory to oscillatory, reminiscent of a spontaneous
breaking of $\mathcal{\mathcal{PT}}$-symmetry \citep{Bender1998,Bender2007,Heiss2012a,El-Ganainy2018,Bender2019,Ashida2020}. We note that, with a small cost in the target state purity, the steering rate can be significantly enhanced. In addition, we present an argument that the appearance of a higher-order exceptional point in the context of optimization is generic and applies to many-body systems as well.

\begin{figure*}
\centering{}\includegraphics[width=1\textwidth]{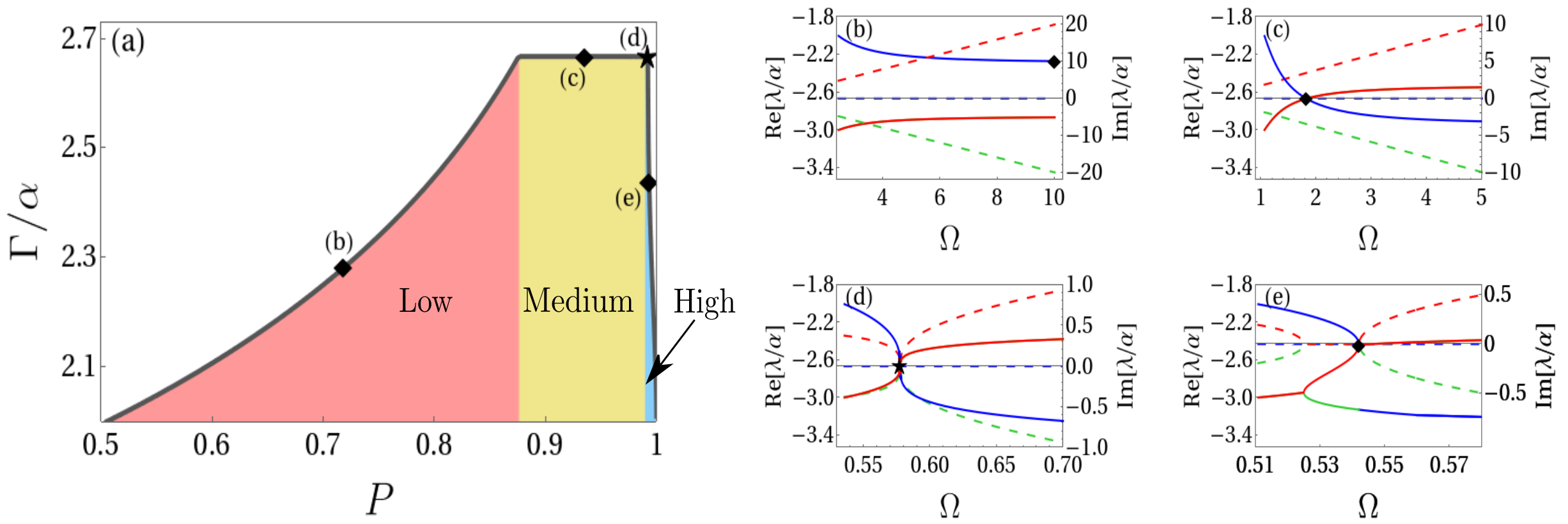}\caption{%
Optimal convergence rate $\Gamma$ in units of the measurement strength $\alpha$ as a function of target state purity $\purity$ (a). We identify three regimes: A low purity regime $1/2 \leq \purity \leq 7/8$, a medium purity regime $7/8 < \purity \leq 127/128$, and a high purity regime $127/128 < \purity \leq 1$.
Choosing one target state from each different purity regime (marked by $\blacklozenge$ in (a)), we plot the real (solid) and imaginary (dashed) parts of the nonzero eigenvalues $\lambda$ of the Liouvillian superoperator as a function
of $\Omega$ in (b), (c) and (e). Here, $\Omega=\omega/\alpha$ is the ratio of the Zeeman energy $\omega$ of the system Hamiltonian, Eq.~\eqref{eq:system Hamiltonian}, to the measurement strength $\alpha$, Eq.~\eqref{eq:master equation}. Note that the eigenvalues $\lambda$ may be real or come in complex-conjugate pairs; in the latter case, the red and the green lines, showing $\mathrm{Re}~\lambda$ of the two complex-conjugate eigenvalues, coincide. Panel (e) contains a segment where all three eigenvalues are real; there all three colors appear explicitly in the plot.
Steering becomes optimal when the parameter $\Omega$ is tuned such that the topmost real part (solid) becomes as negative as possible (marked by $\blacklozenge$).
At the transition from the medium to the high purity (marked by $\star$ in (a)), the optimal convergence rate occurs at a third-order EP shown in (d). \label{fig:steering} }
\end{figure*}

On a more technical level, representing the system state on or within
the Bloch sphere by a vector $\textbf{s}$, its purity is characterized
by $\purity\equiv(1+\textbf{s}^{2})/2$. We find that the optimal steering
dynamics can be characterized by three different regimes, summarized
in Fig.~\ref{fig:steering}. In the low purity regime with $\purity\le7/8$,
the convergence rate becomes optimal by choosing the Zeeman field
in the system Hamiltonian as large as possible, and convergence towards the target state is oscillatory. In the medium
purity regime, $7/8<\purity\le127/128$, optimum convergence
is achieved when all three non-vanishing eigenvalues of the Liouvillian
superoperator have equal real parts; the optimal convergence rate is
independent of $\purity$, while the convergence dynamics
remains oscillatory, similar to the low purity
regime. The transition to the high purity regime with $127/128<\purity\leq 1$
occurs at a third-order EP where all three nonzero eigenvalues of
the Liouvillian superoperator coalesce. In the high purity regime,
the optimal convergence dynamics is non-oscillatory since all
three nonzero eigenvalues of the superoperator are real, and two
of them are degenerate, placing the superoperator at a conventional
second-order EP.

\emph{System evolution and the steady state.---} Consider a two-level
quantum system represented by the density matrix $\rho_{s}$ whose
dynamics comprises two contributions: the unitary evolution and the
measurement evolution. The former  is governed by the following
Hamiltonian acting in the system's Hilbert space
\begin{equation}
H_{s}=\omega\,\hat{n}\cdot\boldsymbol{\sigma},\quad\hat{n}=(\cos\phi\sin\theta,\sin\phi\sin\theta,\cos\theta),\label{eq:system Hamiltonian}
\end{equation}
where $\omega$ is the Zeeman energy of the two levels, $\theta$,
$\phi$ are spherical coordinates parametrizing the unit vector $\hat{n}$,
and $\boldsymbol{\sigma}=(\sigma_{x},\sigma_{y},\sigma_{z})$ is the
vector of Pauli matrices.

For the measurement evolution, the system needs to couple with the
detector, which is chosen to be a two-level quantum object prepared in the
state $\rho_{d}^{0}=\frac{1}{2}(\mathbb{\mathbb{I}}+\hat{m}\cdot\boldsymbol{\sigma})$,
where $\hat{m}$ is the detector state initialization direction. Before
they interact, the joint system-detector state is $\rho(t)=\rho_{s}(t)\otimes\rho_{d}^{0}$.
At later times, the joint state $\rho(t)$ evolves
with the system-detector interaction Hamiltonian
\begin{equation}
H_{s-d}=J\left[\boldsymbol{\sigma}^{s}\cdot\boldsymbol{\sigma}^{d}-(\hat{m}\cdot\boldsymbol{\sigma}^{s})(\hat{m}\cdot\boldsymbol{\sigma}^{d})\right],\label{eq:system-detector Hamiltonian}
\end{equation}
where $J$ is the coupling parameter, and $\boldsymbol{\sigma}^{s}$, $\boldsymbol{\sigma}^{d}$ are the pseudo-spin operators of the system and detector, respectively. Then the interaction is switched
off,  and the detector state is measured projectively; disentangling
the composite system-detector state and generating a measurement back-action
on the system state $\rho_{s}(t)$. In our \emph{blind measurement} protocol~\cite{Roy2019b}, the detector readouts are discarded
(i.e., traced out). After each measurement step, the detector state is reset to $\rho_{d}^{0}$.
We note that the system-detector interaction is chosen to be anisotropic, such that only the
system's spin direction orthogonal to the detector state initialization
direction is coupled.

In our dynamics, the Hamiltonian evolution
(cf.~Eq.~(\ref{eq:system Hamiltonian})) and the measurement evolution
(cf.~Eq.~(\ref{eq:system-detector Hamiltonian})) happen simultaneously.
Over a small time step $dt$, the two processes do not interfere with
each other (up to $O(dt)$).
An infinitesimal time step evolution of the system is given by $\rho_{s}(t+dt)=\textrm{tr}_{\text{d}}[e^{-iHdt}\rho(t)e^{iHdt}]$
where $H=H_{s}\otimes\mathbb{I}+H_{s-d}$.
In the continuous time limit $dt\rightarrow0$,
and using a scaling of $J$ such that $J^{2}dt=\textrm{const}\equiv\alpha$, we obtain~\cite{Roy2019b,Kumar2020}
\begin{equation}
\frac{d\rho_{s}}{dt}=\mathcal{L}[\rho_{s}]=i\,[\rho_{s},H_{s}]-2\alpha\left(L^{\dagger}L\rho_{s}+\rho_{s}L^{\dagger}L-2L\rho_{s}L^{\text{\dag}}\right),\label{eq:master equation}
\end{equation}
where $\mathcal{L}$ is the Liouvillian superoperator, $\alpha$ specifies the measurement strength, and $L=|\hat{m}_{+}\rangle\langle\hat{m}_{-}|$
is the Lindblad jump operator with $|\hat{m}_{\pm}\rangle$ as the
eigenstates of the operator $\hat{m}\cdot\boldsymbol{\sigma}$ with
eigenvalues $\pm1$.
\begin{figure}
\centering{}\includegraphics[width=0.5\textwidth]{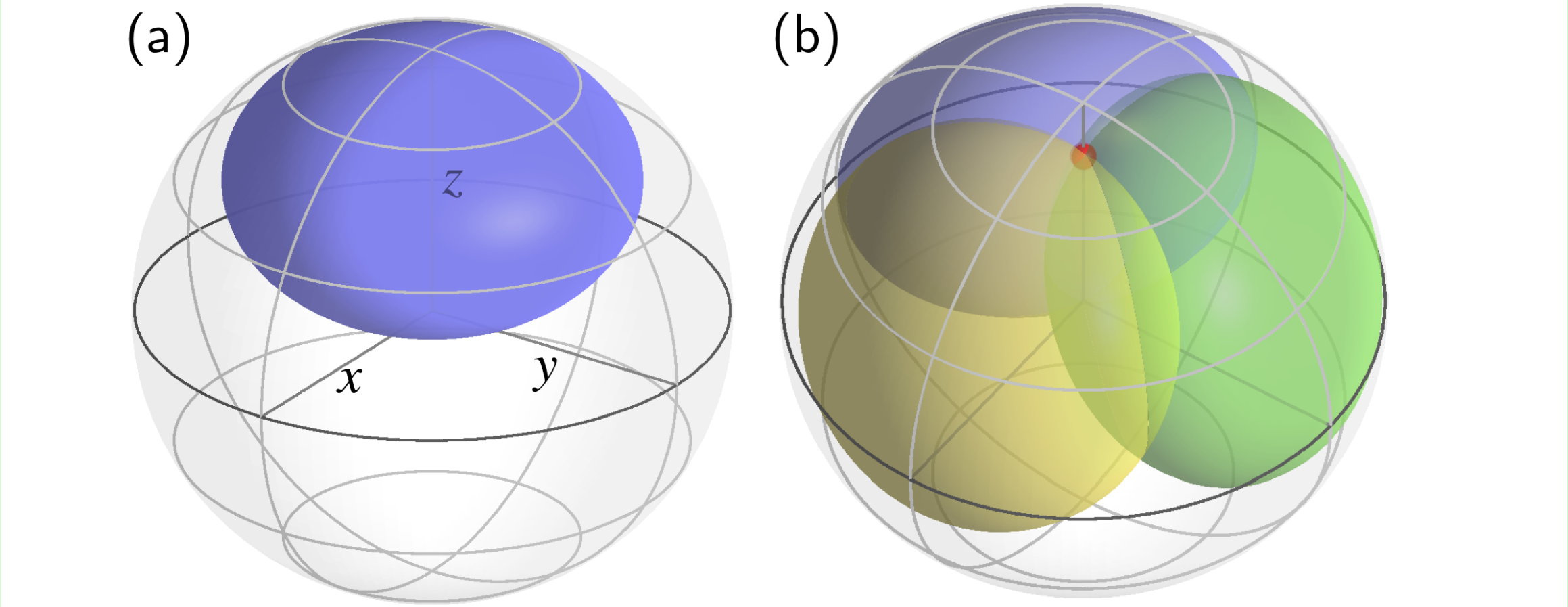}\caption{ (a) Steady state ellipsoid for the detector state initialization
direction $\hat{m}=(0,0,1)^T$, see Eq.~(\ref{eq:steady state coordinates}).
(b) The steady state ellipsoid can be rotated by rotating the detector state initialization direction. Hence, any target state (red) on or inside the Bloch sphere can be reached, as it lies on one or several rotated ellipsoids (e.g. blue, green, yellow).\label{fig:targets}}
\end{figure}

The combined unitary and  weak-measurement  time evolution ultimately
steer the system towards a steady state determined by the condition $d\rho_{s}^{(T)}/dt=\mathcal{L}[\rho_{s}^{(T)}]=0$.
We parameterize the steady state as $\rho_{s}^{(T)}=\frac{1}{2}(\mathbb{I}+\boldsymbol{s}\cdot\boldsymbol{\sigma})$
where $\boldsymbol{s}=(s_{x},s_{y},s_{z})$ is the steady state Bloch
vector. Assuming a detector state initialization $\hat{m}=(0,0,1)$, this steady state is given by
\begin{subequations}
	\label{eq:steady state coordinates}
	\begin{equation}
	s_{x}=\frac{2\Omega\sin\theta(\Omega\cos\theta\cos\phi+\sin\phi)}{2+\Omega^{2}(\cos^{2}\theta+1)},\label{eq: sx steady state coordinate}
	\end{equation}
	\begin{equation}
	s_{y}=\frac{2\Omega\sin\theta(\Omega\cos\theta\sin\phi-\cos\phi)}{2+\Omega^{2}(\cos^{2}\theta+1)},\label{eq:sy steady state coordinate}
	\end{equation}
	\begin{equation}
	s_{z}=\frac{2(1+\Omega^{2}\cos^{2}\theta)}{2+\Omega^{2}(\cos^{2}\theta+1)},\label{eq:sz steady state coordinate}
	\end{equation}
\end{subequations}
where $\Omega=\omega/\alpha$. The steady state coordinates (cf.~Eq.~(\ref{eq:steady state coordinates}))
form an ellipsoid centered at the point $\hat{m}/2$, in this case
$s_{x}^{2}+s_{y}^{2}+2(s_{z}-1/2)^{2}=1/2$; conversely, every state
on this ellipsoid is obtained for at least one choice of the Zeeman field $(\Omega,\theta,\phi)$ 
{[}Fig.~\ref{fig:targets}(a){]}. The main features of the steady
state ellipsoid are as follows: (i) it remains fully confined within
the Bloch sphere, (ii) its shape remains independent of the protocol
parameters, (iii) its minor axis starts from the center of the Bloch
sphere and ends on its surface, coinciding with
the detector state initialization direction $\hat{m}$, implying that
there exists only one pure target state on a given ellipsoid. Each
specific choice of protocol parameters $(\Omega,\theta,\phi)$ steers
the system to a unique steady state on this ellipsoid, but the converse
is not true as a given target state may be stabilized by several distinct sets $(\Omega,\theta,\phi)$. Furthermore, a given steady state may belong to several ellipsoids with different $\hat{m}$, and in that case, the minor axis (which is determined by $\hat{m}$) of
each of these ellipsoids must have a fixed angle with regard to the
Bloch vector $\boldsymbol{s}$ of the steady state (cf.~Fig.~\ref{fig:targets}(b)).

Rotating the detector state initialization direction $\hat{m}$, rotates the ellipsoid. Using all possible $\hat{m}$, the entire Bloch sphere (both surface and interior) can be covered by the ellipsoids, cf.~Fig.~\ref{fig:targets}(b). Therefore, any state, irrespective of its purity, can be targeted using our protocol.

\emph{Optimal steering.---} Our aim now is to optimize the protocol
such that the target state is reached as fast as possible. While the
target state $\rho_{s}^{(T)}$ corresponds to the eigenvector
with  zero eigenvalue of the Liouvillian, $\mathcal{L}[\rho_{s}^{(T)}]=\lambda_{0}\rho_{s}^{(T)}$
with $\lambda_{0}=0$, the dynamical evolution of an arbitrary state
is governed by the eigenvectors $\rho_{s}^{(j)}$ with
nonzero eigenvalues $\lambda_{j}$, i.e.~$\rho_{s}(t)=\rho_{s}^{(T)}+\sum_{j}c_{j}\rho_{s}^{(j)}e^{\lambda_{j}t}$
where the coefficients $c_{j}$ are determined by the system's initial
state $\rho_{s}(0)$. Therefore, the deviations from the target state
decay exponentially in time,  and the decay rates are determined by
the real parts of the nonzero eigenvalues of the superoperator $\mathcal{L}.$
The smallest (in magnitude) nonzero real part, i.e.~the inverse of the Liouvillian
gap, determines the slowest convergence rate ($\Gamma$), and our
aim is to maximize it by choosing appropriate parameters in the protocol.
At first sight, a straightforward way to speed up the steering process
would be to increase the measurement strength $\alpha$, as the Liouvillian
eigenvalues are directly proportional to it.
However, $\alpha = J^2 dt$ is generically limited by the weak measurement constraint
$(J dt)^2 \ll 1$ and by the fact that the experimental measurement and readout time $dt$ cannot be made
arbitrarily short. Therefore, we  consider $\alpha$ to be fixed
at some maximum strength. The protocol parameters that can still
be optimized are the initialization direction $\hat{m}$ and the system
Hamiltonian specified by $\Omega,\,\theta$ and $\phi$. These, however, are partially constrained by the choice of a specific target state. The optimization problem simplifies further by noting that the Liouvillian eigenvalues remain invariant under a unitary rotation of the system state \citep{SupplMat}. Thus, without loss of generality, we analyze the problem by selecting $\hat{m}=(0,0,1)$, for which the steady state is given by Eq.~(\ref{eq:steady state coordinates}) with the corresponding steady state ellipsoid shown in Fig.~\ref{fig:targets}(a). Since different states having the same purity are related via a unitary transformation, all target states on a given ellipsoid having the same purity possess the same convergence rate. This leaves us with two significant parameters: $\theta$ and $\Omega$. We treat $\Omega$ as an independent parameter, while $\theta$ is determined by $\Omega$ and the target state purity $\purity$. Note that, using	the ellipsoid equation, the purity of a steady state  can be expressed as $\purity=1-(1-s_{z})^{2}/2$. For each target state purity $\purity$, we aim to tune the free parameter $\Omega$ such that the convergence rate $\Gamma$ becomes optimal.
However, we note that, for a given target state, there exists a lower
bound on the allowed values of $\Omega$, which is an important constraint on the optimization problem \citep{SupplMat}.

The optimal convergence rate and its dependence on the target state
purity are obtained numerically, and are shown in Fig.~\ref{fig:steering}(a).
We solve the eigenvalue equation $\mathcal{L}[\rho_{s}]=\lambda\,\rho_{s}$
and focus on the eigenvalue having smallest (in magnitude) real part
which we maximize over all admissible values of $\Omega$. We find
that the conditions for optimal convergence  depend on the degree
of purity of the target state --- we identify a low, medium and high
purity regime, Fig.~\ref{fig:steering}(a).
In the low purity regime, the convergence rate becomes bigger the further $\Omega$ is increased, implying that steering becomes optimal by choosing $\Omega$ as large as possible, Fig.~\ref{fig:steering}(b). In the medium purity regime, we find a critical $\Omega$ at which all three nonzero eigenvalues of $\mathcal{L}$ have equal real part, and the convergence rate becomes optimal at this critical $\Omega$, Fig.~\ref{fig:steering}(c).
The transition from the medium to the high purity regime is marked by the fact that at this critical $\Omega$, not only the real parts, but also the imaginary parts of the three nonzero eigenvalues of $\mathcal{L}$ coincide --- the convergence rate becomes optimal at a third-order EP where all three nonzero eigenvalues of $\mathcal{L}$ coincide, Fig.~\ref{fig:steering}(d).
In the high purity regime, we find optimal convergence for a critical $\Omega$ where $\mathcal{L}$ encounters a second-order EP, Fig.~\ref{fig:steering}(e). Intuitively, one can understand the optimized convergence rate behavior in the following way. The protocol targeting any mixed state (with purity $P$ less than 1) involves an interplay of the relaxation to a pure state (controlled by $\alpha$) and Hamiltonian rotation (controlled by $\omega$). Relaxation happening probabilistically at different moments in time, combined with the Hamiltonian rotation of the newly-relaxed states, leads to the desired mixed state. For target states with sufficiently large purities, there is an optimal ratio $\Omega$ = $\omega/\alpha$, when the convergence is the fastest. Whereas for sufficiently small purities, the approach to the target state is limited by the speed of “mixing”, so that it is the faster, the larger $\Omega$ is. The optimal approach to the target state is oscillatory in the low and medium purity regimes, as two of the eigenvalues have nonzero imaginary parts for all values of $\Omega$, Fig.~\ref{fig:steering}(b,c).
By contrast, the optimal approach to the target state becomes non-oscillatory
(exponential decay) in the high purity regime, because all three eigenvalues of the Liouvillian are real at the optimal value of $\Omega$, Fig.~\ref{fig:steering}(e).
Analytical results for the transitions between regimes and the optimal convergence rates are presented in the supplemental material~\citep{SupplMat}.

We highlight that the central feature of our system is a dynamical phase transition at the target state purity $\purity=127/128$ from Hamiltonian-dominated, oscillatory to measurement-dominated, non-oscillatory dynamics. This transition proceeds through a \emph{third-order} EP where all three nonzero eigenvalues of the Liouvillian superoperator coincide; the eigenvalues are obtained as~\citep{SupplMat}: $\lambda_1=\lambda_2=\lambda_3= -8\alpha/3.$
While dynamical phase transitions have been associated with second-order EPs, for instance in $\mathcal{PT}$-symmetric systems~\cite{Heiss2012a,El-Ganainy2018,Ozdemir2019}, the natural appearance of a third-order EP seems striking.

To understand why optimal steering is related to a third-order EP, we now present a general principle for optimization. To explain this principle, we first shift our perspective and ask: Comparing the target states in different regimes, what is the fastest possible convergence rate that can be achieved? It can be shown that, in general, the \textit{average} of the decay rates (i.e., the trace of the Liouvillian superoperator) depends only on the dissipative channels and not on the system Hamiltonian \citep{HeinrichQuote}. In our protocol we fix the average to be $8\alpha/3$. Evidently, if one decay rate is above average, then another one is necessarily below average; therefore, the optimum convergence rate is achieved when all decay rates are equal (to this average) which naturally happens at the EPs. In our protocol this is realized in the medium purity regime (cf. Fig~\ref{fig:steering}(a)). In fact, in this regime the convergence rate plateaus at an upper limit that is given by the \textit{average} of the decay rates. In the other regimes, there is no value of $\Omega$ for which all three decay rates become equal simultaneously. We thus conclude that steering is necessarily optimal at the third-order EP, because the equality of all three nonzero eigenvalues entails the equality of the decay rates.

One expects the above considerations to be applicable to broad classes of systems, including dissipative many-body platforms \citep{HeinrichQuote}. The presence of additional (unwanted) dissipative channels beyond those required by the protocol can modify the optimal steering rate. However, we still expect the optimality to be associated with an exceptional point.

\emph{Experimental implementation.---}Our protocol can be implemented
in a variety of experimental platforms. The main ingredient of our
protocol, blind measurements stabilizing the system at a specific pure
state, is particularly natural for implementation in cold atomic systems
such as cold ions~\citep{BenAv2020} or Rydberg atoms~\citep{Weimer2010}, but is also supported by fluxonium~\citep{Pop2014,Earnest2018,Hazard2019} and transmon qubits~\citep{Chen2021}.
High-fidelity coherent Hamiltonian manipulation in these systems is
now routinely performed in many laboratories \citep{Saffman2010,Madjarov2020,Gan2020}.
Combining the ingredients and implementing our protocol in these systems
appears straightforward. Checking our predictions for the steering
optimality and its relation to EPs, however, will require a degree
of control and stability in the system beyond the standard levels,
as the system sensitivity to perturbations in the vicinity of EPs
is expected to be enhanced \citep{Wiersig2014,Chen2017,Hodaei2017,Wang2020}.

\emph{Conclusion.---} We have proposed a steering protocol
which uses the interplay of a unitary and a weak-measurement-induced evolution to steer a two-level quantum
system towards any desired target state on or within
the Bloch sphere. The resulting Lindbladian dominates the steering towards high purity states, while the Hamiltonian dynamics dominates the optimal steering towards low purity mixed states. In all cases the optimum convergence
rate is linked with exceptional points of the Liouvillian. The latter exhibits a dynamical phase transition such
that the dynamics changes from oscillatory to exponential
decaying. This transition is characterized by
the presence of a third-order EP.

The convergence rate optimization analysis in our work suggests that a significant speedup can be achieved with a slight compromise in the target state purity. This implies that it may be beneficial to target a state that is not exactly the desired one (e.g., has 99\% purity), so that convergence can be sped up significantly. Faster convergence will leave less time for unavoidable noise to affect the system dynamics, thus improving the target state fidelity. This may prove significant for recent state stabilization protocols \citep{Shankar2013,Lin2013,Liu2016,Horn2018}.

\emph{Note added.---} In a recent work \citep{Khandelwal2021b}, Khandelwal et al. have discussed the relevance of EPs for optimal operation of quantum thermal machines.
\begin{acknowledgments}
	We would like to thank Heinrich-Gregor Zirnstein for collaboration on an earlier version of this manuscript. This project was supported by the Deutsche Forschungsgemeinschaft (DFG, German Research Foundation) --- 277101999 --- TRR 183 (project C01), EG 96/13-1 and GO 1405/6-1, by the Israel Science Foundation (ISF), and by the National Science Foundation through award DMR-2037654 and the US--Israel Binational Science Foundation (BSF), Jerusalem, Israel. Y.G. acknowledges support by the Helmholtz International Fellow Award.
\end{acknowledgments}

\bibliography{Suppl,Measurement-induced_state_steering_and_steering_rate_optimization}

\begin{thebibliography}{66}%
\makeatletter
\providecommand \@ifxundefined [1]{%
 \@ifx{#1\undefined}
}%
\providecommand \@ifnum [1]{%
 \ifnum #1\expandafter \@firstoftwo
 \else \expandafter \@secondoftwo
 \fi
}%
\providecommand \@ifx [1]{%
 \ifx #1\expandafter \@firstoftwo
 \else \expandafter \@secondoftwo
 \fi
}%
\providecommand \natexlab [1]{#1}%
\providecommand \enquote  [1]{``#1''}%
\providecommand \bibnamefont  [1]{#1}%
\providecommand \bibfnamefont [1]{#1}%
\providecommand \citenamefont [1]{#1}%
\providecommand \href@noop [0]{\@secondoftwo}%
\providecommand \href [0]{\begingroup \@sanitize@url \@href}%
\providecommand \@href[1]{\@@startlink{#1}\@@href}%
\providecommand \@@href[1]{\endgroup#1\@@endlink}%
\providecommand \@sanitize@url [0]{\catcode `\\12\catcode `\$12\catcode
  `\&12\catcode `\#12\catcode `\^12\catcode `\_12\catcode `\%12\relax}%
\providecommand \@@startlink[1]{}%
\providecommand \@@endlink[0]{}%
\providecommand \url  [0]{\begingroup\@sanitize@url \@url }%
\providecommand \@url [1]{\endgroup\@href {#1}{\urlprefix }}%
\providecommand \urlprefix  [0]{URL }%
\providecommand \Eprint [0]{\href }%
\providecommand \doibase [0]{https://doi.org/}%
\providecommand \selectlanguage [0]{\@gobble}%
\providecommand \bibinfo  [0]{\@secondoftwo}%
\providecommand \bibfield  [0]{\@secondoftwo}%
\providecommand \translation [1]{[#1]}%
\providecommand \BibitemOpen [0]{}%
\providecommand \bibitemStop [0]{}%
\providecommand \bibitemNoStop [0]{.\EOS\space}%
\providecommand \EOS [0]{\spacefactor3000\relax}%
\providecommand \BibitemShut  [1]{\csname bibitem#1\endcsname}%
\let\auto@bib@innerbib\@empty
\bibitem [{\citenamefont {Diehl}\ \emph {et~al.}(2008)\citenamefont {Diehl},
  \citenamefont {Micheli}, \citenamefont {Kantian}, \citenamefont {Kraus},
  \citenamefont {B{\"{u}}chler},\ and\ \citenamefont {Zoller}}]{Diehl2008a}%
  \BibitemOpen
  \bibfield  {author} {\bibinfo {author} {\bibfnamefont {S.}~\bibnamefont
  {Diehl}}, \bibinfo {author} {\bibfnamefont {A.}~\bibnamefont {Micheli}},
  \bibinfo {author} {\bibfnamefont {A.}~\bibnamefont {Kantian}}, \bibinfo
  {author} {\bibfnamefont {B.}~\bibnamefont {Kraus}}, \bibinfo {author}
  {\bibfnamefont {H.~P.}\ \bibnamefont {B{\"{u}}chler}},\ and\ \bibinfo
  {author} {\bibfnamefont {P.}~\bibnamefont {Zoller}},\ }\bibfield  {title}
  {\bibinfo {title} {{Quantum states and phases in driven open quantum systems
  with cold atoms}},\ }\href {https://doi.org/10.1038/nphys1073} {\bibfield
  {journal} {\bibinfo  {journal} {Nature Physics}\ }\textbf {\bibinfo {volume}
  {4}},\ \bibinfo {pages} {878} (\bibinfo {year} {2008})}\BibitemShut {NoStop}%
\bibitem [{\citenamefont {Kraus}\ \emph {et~al.}(2008)\citenamefont {Kraus},
  \citenamefont {B{\"{u}}chler}, \citenamefont {Diehl}, \citenamefont
  {Kantian}, \citenamefont {Micheli},\ and\ \citenamefont
  {Zoller}}]{Kraus2008}%
  \BibitemOpen
  \bibfield  {author} {\bibinfo {author} {\bibfnamefont {B.}~\bibnamefont
  {Kraus}}, \bibinfo {author} {\bibfnamefont {H.~P.}\ \bibnamefont
  {B{\"{u}}chler}}, \bibinfo {author} {\bibfnamefont {S.}~\bibnamefont
  {Diehl}}, \bibinfo {author} {\bibfnamefont {A.}~\bibnamefont {Kantian}},
  \bibinfo {author} {\bibfnamefont {A.}~\bibnamefont {Micheli}},\ and\ \bibinfo
  {author} {\bibfnamefont {P.}~\bibnamefont {Zoller}},\ }\bibfield  {title}
  {\bibinfo {title} {{Preparation of entangled states by quantum Markov
  processes}},\ }\href {https://doi.org/10.1103/PhysRevA.78.042307} {\bibfield
  {journal} {\bibinfo  {journal} {Physical Review A}\ }\textbf {\bibinfo
  {volume} {78}},\ \bibinfo {pages} {042307} (\bibinfo {year}
  {2008})}\BibitemShut {NoStop}%
\bibitem [{\citenamefont {Roncaglia}\ \emph {et~al.}(2010)\citenamefont
  {Roncaglia}, \citenamefont {Rizzi},\ and\ \citenamefont
  {Cirac}}]{Roncaglia2010}%
  \BibitemOpen
  \bibfield  {author} {\bibinfo {author} {\bibfnamefont {M.}~\bibnamefont
  {Roncaglia}}, \bibinfo {author} {\bibfnamefont {M.}~\bibnamefont {Rizzi}},\
  and\ \bibinfo {author} {\bibfnamefont {J.~I.}\ \bibnamefont {Cirac}},\
  }\bibfield  {title} {\bibinfo {title} {{Pfaffian State Generation by Strong
  Three-Body Dissipation}},\ }\href
  {https://doi.org/10.1103/PhysRevLett.104.096803} {\bibfield  {journal}
  {\bibinfo  {journal} {Physical Review Letters}\ }\textbf {\bibinfo {volume}
  {104}},\ \bibinfo {pages} {096803} (\bibinfo {year} {2010})}\BibitemShut
  {NoStop}%
\bibitem [{\citenamefont {Diehl}\ \emph {et~al.}(2011)\citenamefont {Diehl},
  \citenamefont {Rico}, \citenamefont {Baranov},\ and\ \citenamefont
  {Zoller}}]{Diehl2011}%
  \BibitemOpen
  \bibfield  {author} {\bibinfo {author} {\bibfnamefont {S.}~\bibnamefont
  {Diehl}}, \bibinfo {author} {\bibfnamefont {E.}~\bibnamefont {Rico}},
  \bibinfo {author} {\bibfnamefont {M.~A.}\ \bibnamefont {Baranov}},\ and\
  \bibinfo {author} {\bibfnamefont {P.}~\bibnamefont {Zoller}},\ }\bibfield
  {title} {\bibinfo {title} {{Topology by dissipation in atomic quantum
  wires}},\ }\href {https://doi.org/10.1038/nphys2106} {\bibfield  {journal}
  {\bibinfo  {journal} {Nature Physics}\ }\textbf {\bibinfo {volume} {7}},\
  \bibinfo {pages} {971} (\bibinfo {year} {2011})}\BibitemShut {NoStop}%
\bibitem [{\citenamefont {Pechen}(2011)}]{Pechen2011a}%
  \BibitemOpen
  \bibfield  {author} {\bibinfo {author} {\bibfnamefont {A.}~\bibnamefont
  {Pechen}},\ }\bibfield  {title} {\bibinfo {title} {{Engineering arbitrary
  pure and mixed quantum states}},\ }\href
  {https://doi.org/10.1103/PhysRevA.84.042106} {\bibfield  {journal} {\bibinfo
  {journal} {Physical Review A}\ }\textbf {\bibinfo {volume} {84}},\ \bibinfo
  {pages} {042106} (\bibinfo {year} {2011})}\BibitemShut {NoStop}%
\bibitem [{\citenamefont {Murch}\ \emph {et~al.}(2012)\citenamefont {Murch},
  \citenamefont {Vool}, \citenamefont {Zhou}, \citenamefont {Weber},
  \citenamefont {Girvin},\ and\ \citenamefont {Siddiqi}}]{Murch2012}%
  \BibitemOpen
  \bibfield  {author} {\bibinfo {author} {\bibfnamefont {K.~W.}\ \bibnamefont
  {Murch}}, \bibinfo {author} {\bibfnamefont {U.}~\bibnamefont {Vool}},
  \bibinfo {author} {\bibfnamefont {D.}~\bibnamefont {Zhou}}, \bibinfo {author}
  {\bibfnamefont {S.~J.}\ \bibnamefont {Weber}}, \bibinfo {author}
  {\bibfnamefont {S.~M.}\ \bibnamefont {Girvin}},\ and\ \bibinfo {author}
  {\bibfnamefont {I.}~\bibnamefont {Siddiqi}},\ }\bibfield  {title} {\bibinfo
  {title} {{Cavity-Assisted Quantum Bath Engineering}},\ }\href
  {https://doi.org/10.1103/PhysRevLett.109.183602} {\bibfield  {journal}
  {\bibinfo  {journal} {Physical Review Letters}\ }\textbf {\bibinfo {volume}
  {109}},\ \bibinfo {pages} {183602} (\bibinfo {year} {2012})}\BibitemShut
  {NoStop}%
\bibitem [{\citenamefont {Leghtas}\ \emph {et~al.}(2013)\citenamefont
  {Leghtas}, \citenamefont {Vool}, \citenamefont {Shankar}, \citenamefont
  {Hatridge}, \citenamefont {Girvin}, \citenamefont {Devoret},\ and\
  \citenamefont {Mirrahimi}}]{Leghtas2013}%
  \BibitemOpen
  \bibfield  {author} {\bibinfo {author} {\bibfnamefont {Z.}~\bibnamefont
  {Leghtas}}, \bibinfo {author} {\bibfnamefont {U.}~\bibnamefont {Vool}},
  \bibinfo {author} {\bibfnamefont {S.}~\bibnamefont {Shankar}}, \bibinfo
  {author} {\bibfnamefont {M.}~\bibnamefont {Hatridge}}, \bibinfo {author}
  {\bibfnamefont {S.~M.}\ \bibnamefont {Girvin}}, \bibinfo {author}
  {\bibfnamefont {M.~H.}\ \bibnamefont {Devoret}},\ and\ \bibinfo {author}
  {\bibfnamefont {M.}~\bibnamefont {Mirrahimi}},\ }\bibfield  {title} {\bibinfo
  {title} {{Stabilizing a Bell state of two superconducting qubits by
  dissipation engineering}},\ }\href
  {https://doi.org/10.1103/PhysRevA.88.023849} {\bibfield  {journal} {\bibinfo
  {journal} {Physical Review A}\ }\textbf {\bibinfo {volume} {88}},\ \bibinfo
  {pages} {023849} (\bibinfo {year} {2013})}\BibitemShut {NoStop}%
\bibitem [{\citenamefont {Shankar}\ \emph {et~al.}(2013)\citenamefont
  {Shankar}, \citenamefont {Hatridge}, \citenamefont {Leghtas}, \citenamefont
  {Sliwa}, \citenamefont {Narla}, \citenamefont {Vool}, \citenamefont {Girvin},
  \citenamefont {Frunzio}, \citenamefont {Mirrahimi},\ and\ \citenamefont
  {Devoret}}]{Shankar2013}%
  \BibitemOpen
  \bibfield  {author} {\bibinfo {author} {\bibfnamefont {S.}~\bibnamefont
  {Shankar}}, \bibinfo {author} {\bibfnamefont {M.}~\bibnamefont {Hatridge}},
  \bibinfo {author} {\bibfnamefont {Z.}~\bibnamefont {Leghtas}}, \bibinfo
  {author} {\bibfnamefont {K.~M.}\ \bibnamefont {Sliwa}}, \bibinfo {author}
  {\bibfnamefont {A.}~\bibnamefont {Narla}}, \bibinfo {author} {\bibfnamefont
  {U.}~\bibnamefont {Vool}}, \bibinfo {author} {\bibfnamefont {S.~M.}\
  \bibnamefont {Girvin}}, \bibinfo {author} {\bibfnamefont {L.}~\bibnamefont
  {Frunzio}}, \bibinfo {author} {\bibfnamefont {M.}~\bibnamefont {Mirrahimi}},\
  and\ \bibinfo {author} {\bibfnamefont {M.~H.}\ \bibnamefont {Devoret}},\
  }\bibfield  {title} {\bibinfo {title} {{Autonomously stabilized entanglement
  between two superconducting quantum bits}},\ }\href
  {https://doi.org/10.1038/nature12802} {\bibfield  {journal} {\bibinfo
  {journal} {Nature}\ }\textbf {\bibinfo {volume} {504}},\ \bibinfo {pages}
  {419} (\bibinfo {year} {2013})}\BibitemShut {NoStop}%
\bibitem [{\citenamefont {Lin}\ \emph {et~al.}(2013)\citenamefont {Lin},
  \citenamefont {Gaebler}, \citenamefont {Reiter}, \citenamefont {Tan},
  \citenamefont {Bowler}, \citenamefont {S{\o}rensen}, \citenamefont
  {Leibfried},\ and\ \citenamefont {Wineland}}]{Lin2013}%
  \BibitemOpen
  \bibfield  {author} {\bibinfo {author} {\bibfnamefont {Y.}~\bibnamefont
  {Lin}}, \bibinfo {author} {\bibfnamefont {J.~P.}\ \bibnamefont {Gaebler}},
  \bibinfo {author} {\bibfnamefont {F.}~\bibnamefont {Reiter}}, \bibinfo
  {author} {\bibfnamefont {T.~R.}\ \bibnamefont {Tan}}, \bibinfo {author}
  {\bibfnamefont {R.}~\bibnamefont {Bowler}}, \bibinfo {author} {\bibfnamefont
  {A.~S.}\ \bibnamefont {S{\o}rensen}}, \bibinfo {author} {\bibfnamefont
  {D.}~\bibnamefont {Leibfried}},\ and\ \bibinfo {author} {\bibfnamefont
  {D.~J.}\ \bibnamefont {Wineland}},\ }\bibfield  {title} {\bibinfo {title}
  {{Dissipative production of a maximally entangled steady state of two quantum
  bits}},\ }\href {https://doi.org/10.1038/nature12801} {\bibfield  {journal}
  {\bibinfo  {journal} {Nature}\ }\textbf {\bibinfo {volume} {504}},\ \bibinfo
  {pages} {415} (\bibinfo {year} {2013})}\BibitemShut {NoStop}%
\bibitem [{\citenamefont {Liu}\ \emph {et~al.}(2016{\natexlab{a}})\citenamefont
  {Liu}, \citenamefont {Shankar}, \citenamefont {Ofek}, \citenamefont
  {Hatridge}, \citenamefont {Narla}, \citenamefont {Sliwa}, \citenamefont
  {Frunzio}, \citenamefont {Schoelkopf},\ and\ \citenamefont
  {Devoret}}]{Liu2016}%
  \BibitemOpen
  \bibfield  {author} {\bibinfo {author} {\bibfnamefont {Y.}~\bibnamefont
  {Liu}}, \bibinfo {author} {\bibfnamefont {S.}~\bibnamefont {Shankar}},
  \bibinfo {author} {\bibfnamefont {N.}~\bibnamefont {Ofek}}, \bibinfo {author}
  {\bibfnamefont {M.}~\bibnamefont {Hatridge}}, \bibinfo {author}
  {\bibfnamefont {A.}~\bibnamefont {Narla}}, \bibinfo {author} {\bibfnamefont
  {K.~M.}\ \bibnamefont {Sliwa}}, \bibinfo {author} {\bibfnamefont
  {L.}~\bibnamefont {Frunzio}}, \bibinfo {author} {\bibfnamefont {R.~J.}\
  \bibnamefont {Schoelkopf}},\ and\ \bibinfo {author} {\bibfnamefont {M.~H.}\
  \bibnamefont {Devoret}},\ }\bibfield  {title} {\bibinfo {title} {{Comparing
  and Combining Measurement-Based and Driven-Dissipative Entanglement
  Stabilization}},\ }\href {https://doi.org/10.1103/PhysRevX.6.011022}
  {\bibfield  {journal} {\bibinfo  {journal} {Physical Review X}\ }\textbf
  {\bibinfo {volume} {6}},\ \bibinfo {pages} {011022} (\bibinfo {year}
  {2016}{\natexlab{a}})}\BibitemShut {NoStop}%
\bibitem [{\citenamefont {Goldman}\ \emph {et~al.}(2016)\citenamefont
  {Goldman}, \citenamefont {Budich},\ and\ \citenamefont
  {Zoller}}]{Goldman2016}%
  \BibitemOpen
  \bibfield  {author} {\bibinfo {author} {\bibfnamefont {N.}~\bibnamefont
  {Goldman}}, \bibinfo {author} {\bibfnamefont {J.~C.}\ \bibnamefont
  {Budich}},\ and\ \bibinfo {author} {\bibfnamefont {P.}~\bibnamefont
  {Zoller}},\ }\bibfield  {title} {\bibinfo {title} {{Topological quantum
  matter with ultracold gases in optical lattices}},\ }\href
  {https://doi.org/10.1038/nphys3803} {\bibfield  {journal} {\bibinfo
  {journal} {Nature Physics}\ }\textbf {\bibinfo {volume} {12}},\ \bibinfo
  {pages} {639} (\bibinfo {year} {2016})}\BibitemShut {NoStop}%
\bibitem [{\citenamefont {Lu}\ \emph {et~al.}(2017)\citenamefont {Lu},
  \citenamefont {Chakram}, \citenamefont {Leung}, \citenamefont {Earnest},
  \citenamefont {Naik}, \citenamefont {Huang}, \citenamefont {Groszkowski},
  \citenamefont {Kapit}, \citenamefont {Koch},\ and\ \citenamefont
  {Schuster}}]{Lu2017}%
  \BibitemOpen
  \bibfield  {author} {\bibinfo {author} {\bibfnamefont {Y.}~\bibnamefont
  {Lu}}, \bibinfo {author} {\bibfnamefont {S.}~\bibnamefont {Chakram}},
  \bibinfo {author} {\bibfnamefont {N.}~\bibnamefont {Leung}}, \bibinfo
  {author} {\bibfnamefont {N.}~\bibnamefont {Earnest}}, \bibinfo {author}
  {\bibfnamefont {R.~K.}\ \bibnamefont {Naik}}, \bibinfo {author}
  {\bibfnamefont {Z.}~\bibnamefont {Huang}}, \bibinfo {author} {\bibfnamefont
  {P.}~\bibnamefont {Groszkowski}}, \bibinfo {author} {\bibfnamefont
  {E.}~\bibnamefont {Kapit}}, \bibinfo {author} {\bibfnamefont
  {J.}~\bibnamefont {Koch}},\ and\ \bibinfo {author} {\bibfnamefont {D.~I.}\
  \bibnamefont {Schuster}},\ }\bibfield  {title} {\bibinfo {title} {{Universal
  Stabilization of a Parametrically Coupled Qubit}},\ }\href
  {https://doi.org/10.1103/PhysRevLett.119.150502} {\bibfield  {journal}
  {\bibinfo  {journal} {Physical Review Letters}\ }\textbf {\bibinfo {volume}
  {119}},\ \bibinfo {pages} {150502} (\bibinfo {year} {2017})}\BibitemShut
  {NoStop}%
\bibitem [{\citenamefont {Huang}\ \emph {et~al.}(2018)\citenamefont {Huang},
  \citenamefont {Lu}, \citenamefont {Kapit}, \citenamefont {Schuster},\ and\
  \citenamefont {Koch}}]{Huang2018}%
  \BibitemOpen
  \bibfield  {author} {\bibinfo {author} {\bibfnamefont {Z.}~\bibnamefont
  {Huang}}, \bibinfo {author} {\bibfnamefont {Y.}~\bibnamefont {Lu}}, \bibinfo
  {author} {\bibfnamefont {E.}~\bibnamefont {Kapit}}, \bibinfo {author}
  {\bibfnamefont {D.~I.}\ \bibnamefont {Schuster}},\ and\ \bibinfo {author}
  {\bibfnamefont {J.}~\bibnamefont {Koch}},\ }\bibfield  {title} {\bibinfo
  {title} {{Universal stabilization of single-qubit states using a tunable
  coupler}},\ }\href {https://doi.org/10.1103/PhysRevA.97.062345} {\bibfield
  {journal} {\bibinfo  {journal} {Physical Review A}\ }\textbf {\bibinfo
  {volume} {97}},\ \bibinfo {pages} {062345} (\bibinfo {year}
  {2018})}\BibitemShut {NoStop}%
\bibitem [{\citenamefont {Horn}\ \emph {et~al.}(2018)\citenamefont {Horn},
  \citenamefont {Reiter}, \citenamefont {Lin}, \citenamefont {Leibfried},\ and\
  \citenamefont {Koch}}]{Horn2018}%
  \BibitemOpen
  \bibfield  {author} {\bibinfo {author} {\bibfnamefont {K.~P.}\ \bibnamefont
  {Horn}}, \bibinfo {author} {\bibfnamefont {F.}~\bibnamefont {Reiter}},
  \bibinfo {author} {\bibfnamefont {Y.}~\bibnamefont {Lin}}, \bibinfo {author}
  {\bibfnamefont {D.}~\bibnamefont {Leibfried}},\ and\ \bibinfo {author}
  {\bibfnamefont {C.~P.}\ \bibnamefont {Koch}},\ }\bibfield  {title} {\bibinfo
  {title} {{Quantum optimal control of the dissipative production of a
  maximally entangled state}},\ }\href
  {https://doi.org/10.1088/1367-2630/aaf360} {\bibfield  {journal} {\bibinfo
  {journal} {New Journal of Physics}\ }\textbf {\bibinfo {volume} {20}},\
  \bibinfo {pages} {123010} (\bibinfo {year} {2018})}\BibitemShut {NoStop}%
\bibitem [{\citenamefont {Pechen}\ \emph {et~al.}(2006)\citenamefont {Pechen},
  \citenamefont {Il'in}, \citenamefont {Shuang},\ and\ \citenamefont
  {Rabitz}}]{Pechen2006}%
  \BibitemOpen
  \bibfield  {author} {\bibinfo {author} {\bibfnamefont {A.}~\bibnamefont
  {Pechen}}, \bibinfo {author} {\bibfnamefont {N.}~\bibnamefont {Il'in}},
  \bibinfo {author} {\bibfnamefont {F.}~\bibnamefont {Shuang}},\ and\ \bibinfo
  {author} {\bibfnamefont {H.}~\bibnamefont {Rabitz}},\ }\bibfield  {title}
  {\bibinfo {title} {{Quantum control by von Neumann measurements}},\ }\href
  {https://doi.org/10.1103/PhysRevA.74.052102} {\bibfield  {journal} {\bibinfo
  {journal} {Physical Review A}\ }\textbf {\bibinfo {volume} {74}},\ \bibinfo
  {pages} {052102} (\bibinfo {year} {2006})}\BibitemShut {NoStop}%
\bibitem [{\citenamefont {Roa}\ \emph {et~al.}(2006)\citenamefont {Roa},
  \citenamefont {Delgado}, \citenamefont {{Ladr{\'{o}}n de Guevara}},\ and\
  \citenamefont {Klimov}}]{Roa2006}%
  \BibitemOpen
  \bibfield  {author} {\bibinfo {author} {\bibfnamefont {L.}~\bibnamefont
  {Roa}}, \bibinfo {author} {\bibfnamefont {A.}~\bibnamefont {Delgado}},
  \bibinfo {author} {\bibfnamefont {M.~L.}\ \bibnamefont {{Ladr{\'{o}}n de
  Guevara}}},\ and\ \bibinfo {author} {\bibfnamefont {A.~B.}\ \bibnamefont
  {Klimov}},\ }\bibfield  {title} {\bibinfo {title} {{Measurement-driven
  quantum evolution}},\ }\href {https://doi.org/10.1103/PhysRevA.73.012322}
  {\bibfield  {journal} {\bibinfo  {journal} {Physical Review A}\ }\textbf
  {\bibinfo {volume} {73}},\ \bibinfo {pages} {012322} (\bibinfo {year}
  {2006})}\BibitemShut {NoStop}%
\bibitem [{\citenamefont {Roa}\ \emph {et~al.}(2007)\citenamefont {Roa},
  \citenamefont {de~Guevara}, \citenamefont {Delgado}, \citenamefont
  {Olivares-Renter{\'{i}}a},\ and\ \citenamefont {Klimov}}]{Roa2007a}%
  \BibitemOpen
  \bibfield  {author} {\bibinfo {author} {\bibfnamefont {L.}~\bibnamefont
  {Roa}}, \bibinfo {author} {\bibfnamefont {M.~L.~L.}\ \bibnamefont
  {de~Guevara}}, \bibinfo {author} {\bibfnamefont {A.}~\bibnamefont {Delgado}},
  \bibinfo {author} {\bibfnamefont {G.}~\bibnamefont
  {Olivares-Renter{\'{i}}a}},\ and\ \bibinfo {author} {\bibfnamefont {A.~B.}\
  \bibnamefont {Klimov}},\ }\bibfield  {title} {\bibinfo {title} {{Quantum
  evolution by discrete measurements}},\ }\href
  {https://doi.org/10.1088/1742-6596/84/1/012017} {\bibfield  {journal}
  {\bibinfo  {journal} {Journal of Physics: Conference Series}\ }\textbf
  {\bibinfo {volume} {84}},\ \bibinfo {pages} {012017} (\bibinfo {year}
  {2007})}\BibitemShut {NoStop}%
\bibitem [{\citenamefont {Jacobs}(2010)}]{Jacobs2010a}%
  \BibitemOpen
  \bibfield  {author} {\bibinfo {author} {\bibfnamefont {K.}~\bibnamefont
  {Jacobs}},\ }\bibfield  {title} {\bibinfo {title} {{Feedback control using
  only quantum back-action}},\ }\href
  {https://doi.org/10.1088/1367-2630/12/4/043005} {\bibfield  {journal}
  {\bibinfo  {journal} {New Journal of Physics}\ }\textbf {\bibinfo {volume}
  {12}},\ \bibinfo {pages} {043005} (\bibinfo {year} {2010})}\BibitemShut
  {NoStop}%
\bibitem [{\citenamefont {Ashhab}\ and\ \citenamefont
  {Nori}(2010)}]{Ashhab2010}%
  \BibitemOpen
  \bibfield  {author} {\bibinfo {author} {\bibfnamefont {S.}~\bibnamefont
  {Ashhab}}\ and\ \bibinfo {author} {\bibfnamefont {F.}~\bibnamefont {Nori}},\
  }\bibfield  {title} {\bibinfo {title} {{Control-free control: Manipulating a
  quantum system using only a limited set of measurements}},\ }\href
  {https://doi.org/10.1103/PhysRevA.82.062103} {\bibfield  {journal} {\bibinfo
  {journal} {Physical Review A}\ }\textbf {\bibinfo {volume} {82}},\ \bibinfo
  {pages} {062103} (\bibinfo {year} {2010})}\BibitemShut {NoStop}%
\bibitem [{\citenamefont {Roy}\ \emph {et~al.}(2020)\citenamefont {Roy},
  \citenamefont {Chalker}, \citenamefont {Gornyi},\ and\ \citenamefont
  {Gefen}}]{Roy2019b}%
  \BibitemOpen
  \bibfield  {author} {\bibinfo {author} {\bibfnamefont {S.}~\bibnamefont
  {Roy}}, \bibinfo {author} {\bibfnamefont {J.~T.}\ \bibnamefont {Chalker}},
  \bibinfo {author} {\bibfnamefont {I.~V.}\ \bibnamefont {Gornyi}},\ and\
  \bibinfo {author} {\bibfnamefont {Y.}~\bibnamefont {Gefen}},\ }\bibfield
  {title} {\bibinfo {title} {{Measurement-induced steering of quantum
  systems}},\ }\href {https://doi.org/10.1103/PhysRevResearch.2.033347}
  {\bibfield  {journal} {\bibinfo  {journal} {Physical Review Research}\
  }\textbf {\bibinfo {volume} {2}},\ \bibinfo {pages} {033347} (\bibinfo {year}
  {2020})}\BibitemShut {NoStop}%
\bibitem [{\citenamefont {Kumar}\ \emph {et~al.}(2020)\citenamefont {Kumar},
  \citenamefont {Snizhko},\ and\ \citenamefont {Gefen}}]{Kumar2020}%
  \BibitemOpen
  \bibfield  {author} {\bibinfo {author} {\bibfnamefont {P.}~\bibnamefont
  {Kumar}}, \bibinfo {author} {\bibfnamefont {K.}~\bibnamefont {Snizhko}},\
  and\ \bibinfo {author} {\bibfnamefont {Y.}~\bibnamefont {Gefen}},\ }\bibfield
   {title} {\bibinfo {title} {{Engineering two-qubit mixed states with weak
  measurements}},\ }\href {https://doi.org/10.1103/PhysRevResearch.2.042014}
  {\bibfield  {journal} {\bibinfo  {journal} {Physical Review Research}\
  }\textbf {\bibinfo {volume} {2}},\ \bibinfo {pages} {042014} (\bibinfo {year}
  {2020})}\BibitemShut {NoStop}%
\bibitem [{\citenamefont {Ticozzi}\ \emph {et~al.}(2012)\citenamefont
  {Ticozzi}, \citenamefont {Lucchese}, \citenamefont {Cappellaro},\ and\
  \citenamefont {Viola}}]{Ticozzi2012}%
  \BibitemOpen
  \bibfield  {author} {\bibinfo {author} {\bibfnamefont {F.}~\bibnamefont
  {Ticozzi}}, \bibinfo {author} {\bibfnamefont {R.}~\bibnamefont {Lucchese}},
  \bibinfo {author} {\bibfnamefont {P.}~\bibnamefont {Cappellaro}},\ and\
  \bibinfo {author} {\bibfnamefont {L.}~\bibnamefont {Viola}},\ }\bibfield
  {title} {\bibinfo {title} {{Hamiltonian Control of Quantum Dynamical
  Semigroups: Stabilization and Convergence Speed}},\ }\href
  {https://doi.org/10.1109/TAC.2012.2195858} {\bibfield  {journal} {\bibinfo
  {journal} {IEEE Transactions on Automatic Control}\ }\textbf {\bibinfo
  {volume} {57}},\ \bibinfo {pages} {1931} (\bibinfo {year}
  {2012})}\BibitemShut {NoStop}%
\bibitem [{\citenamefont {Baumgartner}\ and\ \citenamefont
  {Narnhofer}(2008)}]{Baumgartner2008}%
  \BibitemOpen
  \bibfield  {author} {\bibinfo {author} {\bibfnamefont {B.}~\bibnamefont
  {Baumgartner}}\ and\ \bibinfo {author} {\bibfnamefont {H.}~\bibnamefont
  {Narnhofer}},\ }\bibfield  {title} {\bibinfo {title} {{Analysis of quantum
  semigroups with GKS–Lindblad generators: II. General}},\ }\href
  {https://doi.org/10.1088/1751-8113/41/39/395303} {\bibfield  {journal}
  {\bibinfo  {journal} {Journal of Physics A: Mathematical and Theoretical}\
  }\textbf {\bibinfo {volume} {41}},\ \bibinfo {pages} {395303} (\bibinfo
  {year} {2008})}\BibitemShut {NoStop}%
\bibitem [{\citenamefont {Kato}(1995)}]{Kato1995}%
  \BibitemOpen
  \bibfield  {author} {\bibinfo {author} {\bibfnamefont {T.}~\bibnamefont
  {Kato}},\ }\href {https://doi.org/10.1007/978-3-642-66282-9} {\emph {\bibinfo
  {title} {{Perturbation Theory for Linear Operators}}}},\ \bibinfo {series}
  {Classics in Mathematics}, Vol.\ \bibinfo {volume} {132}\ (\bibinfo
  {publisher} {Springer Berlin Heidelberg},\ \bibinfo {address} {Berlin,
  Heidelberg},\ \bibinfo {year} {1995})\BibitemShut {NoStop}%
\bibitem [{\citenamefont {Berry}(2004)}]{Berry2004}%
  \BibitemOpen
  \bibfield  {author} {\bibinfo {author} {\bibfnamefont {M.}~\bibnamefont
  {Berry}},\ }\bibfield  {title} {\bibinfo {title} {{Physics of Nonhermitian
  Degeneracies}},\ }\href {https://doi.org/10.1023/B:CJOP.0000044002.05657.04}
  {\bibfield  {journal} {\bibinfo  {journal} {Czechoslovak Journal of Physics}\
  }\textbf {\bibinfo {volume} {54}},\ \bibinfo {pages} {1039} (\bibinfo {year}
  {2004})}\BibitemShut {NoStop}%
\bibitem [{\citenamefont {Heiss}(2012)}]{Heiss2012a}%
  \BibitemOpen
  \bibfield  {author} {\bibinfo {author} {\bibfnamefont {W.~D.}\ \bibnamefont
  {Heiss}},\ }\bibfield  {title} {\bibinfo {title} {{The physics of exceptional
  points}},\ }\href {https://doi.org/10.1088/1751-8113/45/44/444016} {\bibfield
   {journal} {\bibinfo  {journal} {Journal of Physics A: Mathematical and
  Theoretical}\ }\textbf {\bibinfo {volume} {45}},\ \bibinfo {pages} {444016}
  (\bibinfo {year} {2012})}\BibitemShut {NoStop}%
\bibitem [{\citenamefont {Hatano}(2019)}]{Hatano2019a}%
  \BibitemOpen
  \bibfield  {author} {\bibinfo {author} {\bibfnamefont {N.}~\bibnamefont
  {Hatano}},\ }\bibfield  {title} {\bibinfo {title} {{Exceptional points of the
  Lindblad operator of a two-level system}},\ }\href
  {https://doi.org/10.1080/00268976.2019.1593535} {\bibfield  {journal}
  {\bibinfo  {journal} {Molecular Physics}\ }\textbf {\bibinfo {volume}
  {117}},\ \bibinfo {pages} {2121} (\bibinfo {year} {2019})}\BibitemShut
  {NoStop}%
\bibitem [{\citenamefont {Minganti}\ \emph {et~al.}(2019)\citenamefont
  {Minganti}, \citenamefont {Miranowicz}, \citenamefont {Chhajlany},\ and\
  \citenamefont {Nori}}]{Minganti2019}%
  \BibitemOpen
  \bibfield  {author} {\bibinfo {author} {\bibfnamefont {F.}~\bibnamefont
  {Minganti}}, \bibinfo {author} {\bibfnamefont {A.}~\bibnamefont
  {Miranowicz}}, \bibinfo {author} {\bibfnamefont {R.~W.}\ \bibnamefont
  {Chhajlany}},\ and\ \bibinfo {author} {\bibfnamefont {F.}~\bibnamefont
  {Nori}},\ }\bibfield  {title} {\bibinfo {title} {{Quantum exceptional points
  of non-Hermitian Hamiltonians and Liouvillians: The effects of quantum
  jumps}},\ }\href {https://doi.org/10.1103/PhysRevA.100.062131} {\bibfield
  {journal} {\bibinfo  {journal} {Physical Review A}\ }\textbf {\bibinfo
  {volume} {100}},\ \bibinfo {pages} {062131} (\bibinfo {year}
  {2019})}\BibitemShut {NoStop}%
\bibitem [{\citenamefont {Lin}\ \emph {et~al.}(2016)\citenamefont {Lin},
  \citenamefont {Pick}, \citenamefont {Lon{\v{c}}ar},\ and\ \citenamefont
  {Rodriguez}}]{Lin2016}%
  \BibitemOpen
  \bibfield  {author} {\bibinfo {author} {\bibfnamefont {Z.}~\bibnamefont
  {Lin}}, \bibinfo {author} {\bibfnamefont {A.}~\bibnamefont {Pick}}, \bibinfo
  {author} {\bibfnamefont {M.}~\bibnamefont {Lon{\v{c}}ar}},\ and\ \bibinfo
  {author} {\bibfnamefont {A.~W.}\ \bibnamefont {Rodriguez}},\ }\bibfield
  {title} {\bibinfo {title} {{Enhanced Spontaneous Emission at Third-Order
  Dirac Exceptional Points in Inverse-Designed Photonic Crystals}},\ }\href
  {https://doi.org/10.1103/PhysRevLett.117.107402} {\bibfield  {journal}
  {\bibinfo  {journal} {Physical Review Letters}\ }\textbf {\bibinfo {volume}
  {117}},\ \bibinfo {pages} {107402} (\bibinfo {year} {2016})}\BibitemShut
  {NoStop}%
\bibitem [{\citenamefont {Lupu}\ \emph {et~al.}(2017)\citenamefont {Lupu},
  \citenamefont {Konotop},\ and\ \citenamefont {Benisty}}]{Lupu2017}%
  \BibitemOpen
  \bibfield  {author} {\bibinfo {author} {\bibfnamefont {A.}~\bibnamefont
  {Lupu}}, \bibinfo {author} {\bibfnamefont {V.~V.}\ \bibnamefont {Konotop}},\
  and\ \bibinfo {author} {\bibfnamefont {H.}~\bibnamefont {Benisty}},\
  }\bibfield  {title} {\bibinfo {title} {{Optimal PT-symmetric switch features
  exceptional point}},\ }\href {https://doi.org/10.1038/s41598-017-13264-9}
  {\bibfield  {journal} {\bibinfo  {journal} {Scientific Reports}\ }\textbf
  {\bibinfo {volume} {7}},\ \bibinfo {pages} {13299} (\bibinfo {year}
  {2017})}\BibitemShut {NoStop}%
\bibitem [{\citenamefont {Metelmann}\ and\ \citenamefont
  {T{\"{u}}reci}(2018)}]{Metelmann2018}%
  \BibitemOpen
  \bibfield  {author} {\bibinfo {author} {\bibfnamefont {A.}~\bibnamefont
  {Metelmann}}\ and\ \bibinfo {author} {\bibfnamefont {H.~E.}\ \bibnamefont
  {T{\"{u}}reci}},\ }\bibfield  {title} {\bibinfo {title} {{Nonreciprocal
  signal routing in an active quantum network}},\ }\href
  {https://doi.org/10.1103/PhysRevA.97.043833} {\bibfield  {journal} {\bibinfo
  {journal} {Physical Review A}\ }\textbf {\bibinfo {volume} {97}},\ \bibinfo
  {pages} {043833} (\bibinfo {year} {2018})}\BibitemShut {NoStop}%
\bibitem [{\citenamefont {Partanen}\ \emph {et~al.}(2019)\citenamefont
  {Partanen}, \citenamefont {Goetz}, \citenamefont {Tan}, \citenamefont
  {Kohvakka}, \citenamefont {Sevriuk}, \citenamefont {Lake}, \citenamefont
  {Kokkoniemi}, \citenamefont {Ikonen}, \citenamefont {Hazra}, \citenamefont
  {M{\"{a}}kinen}, \citenamefont {Hyypp{\"{a}}}, \citenamefont
  {Gr{\"{o}}nberg}, \citenamefont {Vesterinen}, \citenamefont {Silveri},\ and\
  \citenamefont {M{\"{o}}tt{\"{o}}nen}}]{Partanen2019}%
  \BibitemOpen
  \bibfield  {author} {\bibinfo {author} {\bibfnamefont {M.}~\bibnamefont
  {Partanen}}, \bibinfo {author} {\bibfnamefont {J.}~\bibnamefont {Goetz}},
  \bibinfo {author} {\bibfnamefont {K.~Y.}\ \bibnamefont {Tan}}, \bibinfo
  {author} {\bibfnamefont {K.}~\bibnamefont {Kohvakka}}, \bibinfo {author}
  {\bibfnamefont {V.}~\bibnamefont {Sevriuk}}, \bibinfo {author} {\bibfnamefont
  {R.~E.}\ \bibnamefont {Lake}}, \bibinfo {author} {\bibfnamefont
  {R.}~\bibnamefont {Kokkoniemi}}, \bibinfo {author} {\bibfnamefont
  {J.}~\bibnamefont {Ikonen}}, \bibinfo {author} {\bibfnamefont
  {D.}~\bibnamefont {Hazra}}, \bibinfo {author} {\bibfnamefont
  {A.}~\bibnamefont {M{\"{a}}kinen}}, \bibinfo {author} {\bibfnamefont
  {E.}~\bibnamefont {Hyypp{\"{a}}}}, \bibinfo {author} {\bibfnamefont
  {L.}~\bibnamefont {Gr{\"{o}}nberg}}, \bibinfo {author} {\bibfnamefont
  {V.}~\bibnamefont {Vesterinen}}, \bibinfo {author} {\bibfnamefont
  {M.}~\bibnamefont {Silveri}},\ and\ \bibinfo {author} {\bibfnamefont
  {M.}~\bibnamefont {M{\"{o}}tt{\"{o}}nen}},\ }\bibfield  {title} {\bibinfo
  {title} {{Exceptional points in tunable superconducting resonators}},\ }\href
  {https://doi.org/10.1103/PhysRevB.100.134505} {\bibfield  {journal} {\bibinfo
   {journal} {Physical Review B}\ }\textbf {\bibinfo {volume} {100}},\ \bibinfo
  {pages} {134505} (\bibinfo {year} {2019})}\BibitemShut {NoStop}%
\bibitem [{\citenamefont {Fern{\'{a}}ndez-Alc{\'{a}}zar}\ \emph
  {et~al.}(2020)\citenamefont {Fern{\'{a}}ndez-Alc{\'{a}}zar}, \citenamefont
  {Kononchuk},\ and\ \citenamefont {Kottos}}]{Fernandez-Alcazar2020}%
  \BibitemOpen
  \bibfield  {author} {\bibinfo {author} {\bibfnamefont {L.~J.}\ \bibnamefont
  {Fern{\'{a}}ndez-Alc{\'{a}}zar}}, \bibinfo {author} {\bibfnamefont
  {R.}~\bibnamefont {Kononchuk}},\ and\ \bibinfo {author} {\bibfnamefont
  {T.}~\bibnamefont {Kottos}},\ }\bibfield  {title} {\bibinfo {title} {{Thermal
  Motors with Enhanced Performance due to Engineered Exceptional Points}},\
  }\href {http://arxiv.org/abs/2010.06743} {\  (\bibinfo {year} {2020})},\
  \Eprint {https://arxiv.org/abs/2010.06743} {arXiv:2010.06743} \BibitemShut
  {NoStop}%
\bibitem [{\citenamefont {Wiersig}(2020)}]{Wiersig2020c}%
  \BibitemOpen
  \bibfield  {author} {\bibinfo {author} {\bibfnamefont {J.}~\bibnamefont
  {Wiersig}},\ }\bibfield  {title} {\bibinfo {title} {{Robustness of
  exceptional-point-based sensors against parametric noise: The role of
  Hamiltonian and Liouvillian degeneracies}},\ }\href
  {https://doi.org/10.1103/PhysRevA.101.053846} {\bibfield  {journal} {\bibinfo
   {journal} {Physical Review A}\ }\textbf {\bibinfo {volume} {101}},\ \bibinfo
  {pages} {053846} (\bibinfo {year} {2020})}\BibitemShut {NoStop}%
\bibitem [{\citenamefont {Chen}\ \emph {et~al.}(2021)\citenamefont {Chen},
  \citenamefont {Abbasi}, \citenamefont {Joglekar},\ and\ \citenamefont
  {Murch}}]{Chen2021}%
  \BibitemOpen
  \bibfield  {author} {\bibinfo {author} {\bibfnamefont {W.}~\bibnamefont
  {Chen}}, \bibinfo {author} {\bibfnamefont {M.}~\bibnamefont {Abbasi}},
  \bibinfo {author} {\bibfnamefont {Y.~N.}\ \bibnamefont {Joglekar}},\ and\
  \bibinfo {author} {\bibfnamefont {K.~W.}\ \bibnamefont {Murch}},\ }\bibfield
  {title} {\bibinfo {title} {{Quantum jumps in the non-Hermitian dynamics of a
  superconducting qubit}},\ }\href {http://arxiv.org/abs/2103.06274} {\ \textbf
  {\bibinfo {volume} {2}},\ \bibinfo {pages} {1} (\bibinfo {year} {2021})},\
  \Eprint {https://arxiv.org/abs/2103.06274} {arXiv:2103.06274} \BibitemShut
  {NoStop}%
\bibitem [{\citenamefont {Minganti}\ \emph {et~al.}(2020)\citenamefont
  {Minganti}, \citenamefont {Miranowicz}, \citenamefont {Chhajlany},
  \citenamefont {Arkhipov},\ and\ \citenamefont {Nori}}]{Minganti2020c}%
  \BibitemOpen
  \bibfield  {author} {\bibinfo {author} {\bibfnamefont {F.}~\bibnamefont
  {Minganti}}, \bibinfo {author} {\bibfnamefont {A.}~\bibnamefont
  {Miranowicz}}, \bibinfo {author} {\bibfnamefont {R.~W.}\ \bibnamefont
  {Chhajlany}}, \bibinfo {author} {\bibfnamefont {I.~I.}\ \bibnamefont
  {Arkhipov}},\ and\ \bibinfo {author} {\bibfnamefont {F.}~\bibnamefont
  {Nori}},\ }\bibfield  {title} {\bibinfo {title} {{Hybrid-Liouvillian
  formalism connecting exceptional points of non-Hermitian Hamiltonians and
  Liouvillians via postselection of quantum trajectories}},\ }\href
  {https://doi.org/10.1103/PhysRevA.101.062112} {\bibfield  {journal} {\bibinfo
   {journal} {Physical Review A}\ }\textbf {\bibinfo {volume} {101}},\ \bibinfo
  {pages} {062112} (\bibinfo {year} {2020})}\BibitemShut {NoStop}%
\bibitem [{\citenamefont {Arkhipov}\ \emph {et~al.}(2020)\citenamefont
  {Arkhipov}, \citenamefont {Miranowicz}, \citenamefont {Minganti},\ and\
  \citenamefont {Nori}}]{Arkhipov2020b}%
  \BibitemOpen
  \bibfield  {author} {\bibinfo {author} {\bibfnamefont {I.~I.}\ \bibnamefont
  {Arkhipov}}, \bibinfo {author} {\bibfnamefont {A.}~\bibnamefont
  {Miranowicz}}, \bibinfo {author} {\bibfnamefont {F.}~\bibnamefont
  {Minganti}},\ and\ \bibinfo {author} {\bibfnamefont {F.}~\bibnamefont
  {Nori}},\ }\bibfield  {title} {\bibinfo {title} {{Liouvillian exceptional
  points of any order in dissipative linear bosonic systems: Coherence
  functions and switching between $\mathcal{PT}$ and anti- $\mathcal{PT}$
  symmetries}},\ }\href {https://doi.org/10.1103/PhysRevA.102.033715}
  {\bibfield  {journal} {\bibinfo  {journal} {Physical Review A}\ }\textbf
  {\bibinfo {volume} {102}},\ \bibinfo {pages} {033715} (\bibinfo {year}
  {2020})}\BibitemShut {NoStop}%
\bibitem [{\citenamefont {Lin}\ \emph {et~al.}(2011)\citenamefont {Lin},
  \citenamefont {Ramezani}, \citenamefont {Eichelkraut}, \citenamefont
  {Kottos}, \citenamefont {Cao},\ and\ \citenamefont
  {Christodoulides}}]{Lin2011}%
  \BibitemOpen
  \bibfield  {author} {\bibinfo {author} {\bibfnamefont {Z.}~\bibnamefont
  {Lin}}, \bibinfo {author} {\bibfnamefont {H.}~\bibnamefont {Ramezani}},
  \bibinfo {author} {\bibfnamefont {T.}~\bibnamefont {Eichelkraut}}, \bibinfo
  {author} {\bibfnamefont {T.}~\bibnamefont {Kottos}}, \bibinfo {author}
  {\bibfnamefont {H.}~\bibnamefont {Cao}},\ and\ \bibinfo {author}
  {\bibfnamefont {D.~N.}\ \bibnamefont {Christodoulides}},\ }\bibfield  {title}
  {\bibinfo {title} {{Unidirectional Invisibility Induced by PT -Symmetric
  Periodic Structures}},\ }\href
  {https://doi.org/10.1103/PhysRevLett.106.213901} {\bibfield  {journal}
  {\bibinfo  {journal} {Physical Review Letters}\ }\textbf {\bibinfo {volume}
  {106}},\ \bibinfo {pages} {213901} (\bibinfo {year} {2011})}\BibitemShut
  {NoStop}%
\bibitem [{\citenamefont {Regensburger}\ \emph {et~al.}(2012)\citenamefont
  {Regensburger}, \citenamefont {Bersch}, \citenamefont {Miri}, \citenamefont
  {Onishchukov}, \citenamefont {Christodoulides},\ and\ \citenamefont
  {Peschel}}]{Regensburger2012}%
  \BibitemOpen
  \bibfield  {author} {\bibinfo {author} {\bibfnamefont {A.}~\bibnamefont
  {Regensburger}}, \bibinfo {author} {\bibfnamefont {C.}~\bibnamefont
  {Bersch}}, \bibinfo {author} {\bibfnamefont {M.-A.}\ \bibnamefont {Miri}},
  \bibinfo {author} {\bibfnamefont {G.}~\bibnamefont {Onishchukov}}, \bibinfo
  {author} {\bibfnamefont {D.~N.}\ \bibnamefont {Christodoulides}},\ and\
  \bibinfo {author} {\bibfnamefont {U.}~\bibnamefont {Peschel}},\ }\bibfield
  {title} {\bibinfo {title} {{Parity–time synthetic photonic lattices}},\
  }\href {https://doi.org/10.1038/nature11298} {\bibfield  {journal} {\bibinfo
  {journal} {Nature}\ }\textbf {\bibinfo {volume} {488}},\ \bibinfo {pages}
  {167} (\bibinfo {year} {2012})}\BibitemShut {NoStop}%
\bibitem [{\citenamefont {Feng}\ \emph {et~al.}(2013)\citenamefont {Feng},
  \citenamefont {Xu}, \citenamefont {Fegadolli}, \citenamefont {Lu},
  \citenamefont {Oliveira}, \citenamefont {Almeida}, \citenamefont {Chen},\
  and\ \citenamefont {Scherer}}]{Feng2013}%
  \BibitemOpen
  \bibfield  {author} {\bibinfo {author} {\bibfnamefont {L.}~\bibnamefont
  {Feng}}, \bibinfo {author} {\bibfnamefont {Y.-L.}\ \bibnamefont {Xu}},
  \bibinfo {author} {\bibfnamefont {W.~S.}\ \bibnamefont {Fegadolli}}, \bibinfo
  {author} {\bibfnamefont {M.-H.}\ \bibnamefont {Lu}}, \bibinfo {author}
  {\bibfnamefont {J.~E.~B.}\ \bibnamefont {Oliveira}}, \bibinfo {author}
  {\bibfnamefont {V.~R.}\ \bibnamefont {Almeida}}, \bibinfo {author}
  {\bibfnamefont {Y.-F.}\ \bibnamefont {Chen}},\ and\ \bibinfo {author}
  {\bibfnamefont {A.}~\bibnamefont {Scherer}},\ }\bibfield  {title} {\bibinfo
  {title} {{Experimental demonstration of a unidirectional reflectionless
  parity-time metamaterial at optical frequencies}},\ }\href
  {https://doi.org/10.1038/nmat3495} {\bibfield  {journal} {\bibinfo  {journal}
  {Nature Materials}\ }\textbf {\bibinfo {volume} {12}},\ \bibinfo {pages}
  {108} (\bibinfo {year} {2013})}\BibitemShut {NoStop}%
\bibitem [{\citenamefont {Peng}\ \emph {et~al.}(2014)\citenamefont {Peng},
  \citenamefont {{\"{O}}zdemir}, \citenamefont {Lei}, \citenamefont {Monifi},
  \citenamefont {Gianfreda}, \citenamefont {Long}, \citenamefont {Fan},
  \citenamefont {Nori}, \citenamefont {Bender},\ and\ \citenamefont
  {Yang}}]{Peng2014}%
  \BibitemOpen
  \bibfield  {author} {\bibinfo {author} {\bibfnamefont {B.}~\bibnamefont
  {Peng}}, \bibinfo {author} {\bibfnamefont {a.~K.}\ \bibnamefont
  {{\"{O}}zdemir}}, \bibinfo {author} {\bibfnamefont {F.}~\bibnamefont {Lei}},
  \bibinfo {author} {\bibfnamefont {F.}~\bibnamefont {Monifi}}, \bibinfo
  {author} {\bibfnamefont {M.}~\bibnamefont {Gianfreda}}, \bibinfo {author}
  {\bibfnamefont {G.~L.}\ \bibnamefont {Long}}, \bibinfo {author}
  {\bibfnamefont {S.}~\bibnamefont {Fan}}, \bibinfo {author} {\bibfnamefont
  {F.}~\bibnamefont {Nori}}, \bibinfo {author} {\bibfnamefont {C.~M.}\
  \bibnamefont {Bender}},\ and\ \bibinfo {author} {\bibfnamefont
  {L.}~\bibnamefont {Yang}},\ }\bibfield  {title} {\bibinfo {title}
  {{Parity–time-symmetric whispering-gallery microcavities}},\ }\href
  {https://doi.org/10.1038/nphys2927} {\bibfield  {journal} {\bibinfo
  {journal} {Nature Physics}\ }\textbf {\bibinfo {volume} {10}},\ \bibinfo
  {pages} {394} (\bibinfo {year} {2014})}\BibitemShut {NoStop}%
\bibitem [{\citenamefont {Wiersig}(2014)}]{Wiersig2014}%
  \BibitemOpen
  \bibfield  {author} {\bibinfo {author} {\bibfnamefont {J.}~\bibnamefont
  {Wiersig}},\ }\bibfield  {title} {\bibinfo {title} {{Enhancing the
  Sensitivity of Frequency and Energy Splitting Detection by Using Exceptional
  Points: Application to Microcavity Sensors for Single-Particle Detection}},\
  }\href {https://doi.org/10.1103/PhysRevLett.112.203901} {\bibfield  {journal}
  {\bibinfo  {journal} {Physical Review Letters}\ }\textbf {\bibinfo {volume}
  {112}},\ \bibinfo {pages} {203901} (\bibinfo {year} {2014})}\BibitemShut
  {NoStop}%
\bibitem [{\citenamefont {Liu}\ \emph {et~al.}(2016{\natexlab{b}})\citenamefont
  {Liu}, \citenamefont {Zhang}, \citenamefont {{\"{O}}zdemir}, \citenamefont
  {Peng}, \citenamefont {Jing}, \citenamefont {L{\"{u}}}, \citenamefont {Li},
  \citenamefont {Yang}, \citenamefont {Nori},\ and\ \citenamefont
  {Liu}}]{Liu2016a}%
  \BibitemOpen
  \bibfield  {author} {\bibinfo {author} {\bibfnamefont {Z.-P.}\ \bibnamefont
  {Liu}}, \bibinfo {author} {\bibfnamefont {J.}~\bibnamefont {Zhang}}, \bibinfo
  {author} {\bibfnamefont {a.~K.}\ \bibnamefont {{\"{O}}zdemir}}, \bibinfo
  {author} {\bibfnamefont {B.}~\bibnamefont {Peng}}, \bibinfo {author}
  {\bibfnamefont {H.}~\bibnamefont {Jing}}, \bibinfo {author} {\bibfnamefont
  {X.-Y.}\ \bibnamefont {L{\"{u}}}}, \bibinfo {author} {\bibfnamefont {C.-W.}\
  \bibnamefont {Li}}, \bibinfo {author} {\bibfnamefont {L.}~\bibnamefont
  {Yang}}, \bibinfo {author} {\bibfnamefont {F.}~\bibnamefont {Nori}},\ and\
  \bibinfo {author} {\bibfnamefont {Y.-x.}\ \bibnamefont {Liu}},\ }\bibfield
  {title} {\bibinfo {title} {{Metrology with PT-Symmetric Cavities: Enhanced
  Sensitivity near the PT-Phase Transition}},\ }\href
  {https://doi.org/10.1103/PhysRevLett.117.110802} {\bibfield  {journal}
  {\bibinfo  {journal} {Physical Review Letters}\ }\textbf {\bibinfo {volume}
  {117}},\ \bibinfo {pages} {110802} (\bibinfo {year}
  {2016}{\natexlab{b}})}\BibitemShut {NoStop}%
\bibitem [{\citenamefont {Hodaei}\ \emph {et~al.}(2017)\citenamefont {Hodaei},
  \citenamefont {Hassan}, \citenamefont {Wittek}, \citenamefont
  {Garcia-Gracia}, \citenamefont {El-Ganainy}, \citenamefont
  {Christodoulides},\ and\ \citenamefont {Khajavikhan}}]{Hodaei2017}%
  \BibitemOpen
  \bibfield  {author} {\bibinfo {author} {\bibfnamefont {H.}~\bibnamefont
  {Hodaei}}, \bibinfo {author} {\bibfnamefont {A.~U.}\ \bibnamefont {Hassan}},
  \bibinfo {author} {\bibfnamefont {S.}~\bibnamefont {Wittek}}, \bibinfo
  {author} {\bibfnamefont {H.}~\bibnamefont {Garcia-Gracia}}, \bibinfo {author}
  {\bibfnamefont {R.}~\bibnamefont {El-Ganainy}}, \bibinfo {author}
  {\bibfnamefont {D.~N.}\ \bibnamefont {Christodoulides}},\ and\ \bibinfo
  {author} {\bibfnamefont {M.}~\bibnamefont {Khajavikhan}},\ }\bibfield
  {title} {\bibinfo {title} {{Enhanced sensitivity at higher-order exceptional
  points}},\ }\href {https://doi.org/10.1038/nature23280} {\bibfield  {journal}
  {\bibinfo  {journal} {Nature}\ }\textbf {\bibinfo {volume} {548}},\ \bibinfo
  {pages} {187} (\bibinfo {year} {2017})}\BibitemShut {NoStop}%
\bibitem [{\citenamefont {Chen}\ \emph {et~al.}(2017)\citenamefont {Chen},
  \citenamefont {{Kaya {\"{O}}zdemir}}, \citenamefont {Zhao}, \citenamefont
  {Wiersig},\ and\ \citenamefont {Yang}}]{Chen2017}%
  \BibitemOpen
  \bibfield  {author} {\bibinfo {author} {\bibfnamefont {W.}~\bibnamefont
  {Chen}}, \bibinfo {author} {\bibfnamefont {a.}~\bibnamefont {{Kaya
  {\"{O}}zdemir}}}, \bibinfo {author} {\bibfnamefont {G.}~\bibnamefont {Zhao}},
  \bibinfo {author} {\bibfnamefont {J.}~\bibnamefont {Wiersig}},\ and\ \bibinfo
  {author} {\bibfnamefont {L.}~\bibnamefont {Yang}},\ }\bibfield  {title}
  {\bibinfo {title} {{Exceptional points enhance sensing in an optical
  microcavity}},\ }\href {https://doi.org/10.1038/nature23281} {\bibfield
  {journal} {\bibinfo  {journal} {Nature}\ }\textbf {\bibinfo {volume} {548}},\
  \bibinfo {pages} {192} (\bibinfo {year} {2017})}\BibitemShut {NoStop}%
\bibitem [{\citenamefont {Lau}\ and\ \citenamefont {Clerk}(2018)}]{Lau2018}%
  \BibitemOpen
  \bibfield  {author} {\bibinfo {author} {\bibfnamefont {H.-K.}\ \bibnamefont
  {Lau}}\ and\ \bibinfo {author} {\bibfnamefont {A.~A.}\ \bibnamefont
  {Clerk}},\ }\bibfield  {title} {\bibinfo {title} {{Fundamental limits and
  non-reciprocal approaches in non-Hermitian quantum sensing}},\ }\href
  {https://doi.org/10.1038/s41467-018-06477-7} {\bibfield  {journal} {\bibinfo
  {journal} {Nature Communications}\ }\textbf {\bibinfo {volume} {9}},\
  \bibinfo {pages} {4320} (\bibinfo {year} {2018})}\BibitemShut {NoStop}%
\bibitem [{\citenamefont {Zhang}\ \emph {et~al.}(2019)\citenamefont {Zhang},
  \citenamefont {Sweeney}, \citenamefont {Hsu}, \citenamefont {Yang},
  \citenamefont {Stone},\ and\ \citenamefont {Jiang}}]{Zhang2019b}%
  \BibitemOpen
  \bibfield  {author} {\bibinfo {author} {\bibfnamefont {M.}~\bibnamefont
  {Zhang}}, \bibinfo {author} {\bibfnamefont {W.}~\bibnamefont {Sweeney}},
  \bibinfo {author} {\bibfnamefont {C.~W.}\ \bibnamefont {Hsu}}, \bibinfo
  {author} {\bibfnamefont {L.}~\bibnamefont {Yang}}, \bibinfo {author}
  {\bibfnamefont {A.~D.}\ \bibnamefont {Stone}},\ and\ \bibinfo {author}
  {\bibfnamefont {L.}~\bibnamefont {Jiang}},\ }\bibfield  {title} {\bibinfo
  {title} {{Quantum Noise Theory of Exceptional Point Amplifying Sensors}},\
  }\href {https://doi.org/10.1103/PhysRevLett.123.180501} {\bibfield  {journal}
  {\bibinfo  {journal} {Physical Review Letters}\ }\textbf {\bibinfo {volume}
  {123}},\ \bibinfo {pages} {180501} (\bibinfo {year} {2019})}\BibitemShut
  {NoStop}%
\bibitem [{Note1()}]{Note1}%
  \BibitemOpen
  \bibinfo {note} {In this paper, we use the word ``steering'' as the name of a
  process that leads an arbitrary initial state of the system to a
  predesignated target state. This should not be confused with ``steering''
  from the theory of quantum information and quantum computation which defines
  a special kind of nonlocal correlations.}\BibitemShut {Stop}%
\bibitem [{\citenamefont {Bender}\ and\ \citenamefont
  {Boettcher}(1998)}]{Bender1998}%
  \BibitemOpen
  \bibfield  {author} {\bibinfo {author} {\bibfnamefont {C.~M.}\ \bibnamefont
  {Bender}}\ and\ \bibinfo {author} {\bibfnamefont {S.}~\bibnamefont
  {Boettcher}},\ }\bibfield  {title} {\bibinfo {title} {{Real Spectra in
  Non-Hermitian Hamiltonians Having PT Symmetry}},\ }\href
  {https://doi.org/10.1103/PhysRevLett.80.5243} {\bibfield  {journal} {\bibinfo
   {journal} {Physical Review Letters}\ }\textbf {\bibinfo {volume} {80}},\
  \bibinfo {pages} {5243} (\bibinfo {year} {1998})}\BibitemShut {NoStop}%
\bibitem [{\citenamefont {Bender}(2007)}]{Bender2007}%
  \BibitemOpen
  \bibfield  {author} {\bibinfo {author} {\bibfnamefont {C.~M.}\ \bibnamefont
  {Bender}},\ }\bibfield  {title} {\bibinfo {title} {{Making sense of
  non-Hermitian Hamiltonians}},\ }\href
  {https://doi.org/10.1088/0034-4885/70/6/R03} {\bibfield  {journal} {\bibinfo
  {journal} {Reports on Progress in Physics}\ }\textbf {\bibinfo {volume}
  {70}},\ \bibinfo {pages} {947} (\bibinfo {year} {2007})}\BibitemShut
  {NoStop}%
\bibitem [{\citenamefont {El-Ganainy}\ \emph {et~al.}(2018)\citenamefont
  {El-Ganainy}, \citenamefont {Makris}, \citenamefont {Khajavikhan},
  \citenamefont {Musslimani}, \citenamefont {Rotter},\ and\ \citenamefont
  {Christodoulides}}]{El-Ganainy2018}%
  \BibitemOpen
  \bibfield  {author} {\bibinfo {author} {\bibfnamefont {R.}~\bibnamefont
  {El-Ganainy}}, \bibinfo {author} {\bibfnamefont {K.~G.}\ \bibnamefont
  {Makris}}, \bibinfo {author} {\bibfnamefont {M.}~\bibnamefont {Khajavikhan}},
  \bibinfo {author} {\bibfnamefont {Z.~H.}\ \bibnamefont {Musslimani}},
  \bibinfo {author} {\bibfnamefont {S.}~\bibnamefont {Rotter}},\ and\ \bibinfo
  {author} {\bibfnamefont {D.~N.}\ \bibnamefont {Christodoulides}},\ }\bibfield
   {title} {\bibinfo {title} {{Non-Hermitian physics and PT symmetry}},\ }\href
  {https://doi.org/10.1038/nphys4323} {\bibfield  {journal} {\bibinfo
  {journal} {Nature Physics}\ }\textbf {\bibinfo {volume} {14}},\ \bibinfo
  {pages} {11} (\bibinfo {year} {2018})}\BibitemShut {NoStop}%
\bibitem [{\citenamefont {Bender}\ \emph {et~al.}(2019)\citenamefont {Bender},
  \citenamefont {Dorey}, \citenamefont {Dunning}, \citenamefont {Fring},
  \citenamefont {Hook}, \citenamefont {Jones}, \citenamefont {Kuzhel},
  \citenamefont {L{\'{e}}vai},\ and\ \citenamefont {Tateo}}]{Bender2019}%
  \BibitemOpen
  \bibfield  {author} {\bibinfo {author} {\bibfnamefont {C.~M.}\ \bibnamefont
  {Bender}}, \bibinfo {author} {\bibfnamefont {P.~E.}\ \bibnamefont {Dorey}},
  \bibinfo {author} {\bibfnamefont {C.}~\bibnamefont {Dunning}}, \bibinfo
  {author} {\bibfnamefont {A.}~\bibnamefont {Fring}}, \bibinfo {author}
  {\bibfnamefont {D.~W.}\ \bibnamefont {Hook}}, \bibinfo {author}
  {\bibfnamefont {H.~F.}\ \bibnamefont {Jones}}, \bibinfo {author}
  {\bibfnamefont {S.}~\bibnamefont {Kuzhel}}, \bibinfo {author} {\bibfnamefont
  {G.}~\bibnamefont {L{\'{e}}vai}},\ and\ \bibinfo {author} {\bibfnamefont
  {R.}~\bibnamefont {Tateo}},\ }\href {https://doi.org/10.1142/q0178} {\emph
  {\bibinfo {title} {PT Symmetry: In Quantum and Classical Physics}}}\
  (\bibinfo  {publisher} {World Scientific (Europe)},\ \bibinfo {year} {2019})\
  Chap.~\bibinfo {chapter} {1}\BibitemShut {NoStop}%
\bibitem [{\citenamefont {Ashida}\ \emph {et~al.}(2020)\citenamefont {Ashida},
  \citenamefont {Gong},\ and\ \citenamefont {Ueda}}]{Ashida2020}%
  \BibitemOpen
  \bibfield  {author} {\bibinfo {author} {\bibfnamefont {Y.}~\bibnamefont
  {Ashida}}, \bibinfo {author} {\bibfnamefont {Z.}~\bibnamefont {Gong}},\ and\
  \bibinfo {author} {\bibfnamefont {M.}~\bibnamefont {Ueda}},\ }\bibfield
  {title} {\bibinfo {title} {{Non-Hermitian Physics}},\ }\href
  {http://arxiv.org/abs/2006.01837} {\bibfield  {journal} {\bibinfo  {journal}
  {arXiv}\ } (\bibinfo {year} {2020})},\ \Eprint
  {https://arxiv.org/abs/2006.01837} {arXiv:2006.01837} \BibitemShut {NoStop}%
\bibitem [{Sup()}]{SupplMat}%
  \BibitemOpen
  \href@noop {} {}\bibinfo {note} {See the Supplemental Material where we
  derive the lower bound on allowed $\Omega$ and obtain the optimal convergence
  rate.}\BibitemShut {Stop}%
\bibitem [{\citenamefont {{\"O}zdemir}\ \emph {et~al.}(2019)\citenamefont
  {{\"O}zdemir}, \citenamefont {Rotter}, \citenamefont {Nori},\ and\
  \citenamefont {Yang}}]{Ozdemir2019}%
  \BibitemOpen
  \bibfield  {author} {\bibinfo {author} {\bibfnamefont {{\c S}.~K.}\
  \bibnamefont {{\"O}zdemir}}, \bibinfo {author} {\bibfnamefont
  {S.}~\bibnamefont {Rotter}}, \bibinfo {author} {\bibfnamefont
  {F.}~\bibnamefont {Nori}},\ and\ \bibinfo {author} {\bibfnamefont
  {L.}~\bibnamefont {Yang}},\ }\bibfield  {title} {\bibinfo {title}
  {Parity-time symmetry and exceptional points in photonics},\ }\href
  {https://doi.org/10.1038/s41563-019-0304-9} {\bibfield  {journal} {\bibinfo
  {journal} {Nat. Mater.}\ }\textbf {\bibinfo {volume} {18}},\ \bibinfo {pages}
  {783} (\bibinfo {year} {2019})}\BibitemShut {NoStop}%
\bibitem [{Hei()}]{HeinrichQuote}%
  \BibitemOpen
  \href@noop {} {}\bibinfo {note} {H.-G. Zirnstein, unpublished.}\BibitemShut
  {Stop}%
\bibitem [{\citenamefont {{Ben Av}}\ \emph {et~al.}(2020)\citenamefont {{Ben
  Av}}, \citenamefont {Shapira}, \citenamefont {Akerman},\ and\ \citenamefont
  {Ozeri}}]{BenAv2020}%
  \BibitemOpen
  \bibfield  {author} {\bibinfo {author} {\bibfnamefont {E.}~\bibnamefont {{Ben
  Av}}}, \bibinfo {author} {\bibfnamefont {Y.}~\bibnamefont {Shapira}},
  \bibinfo {author} {\bibfnamefont {N.}~\bibnamefont {Akerman}},\ and\ \bibinfo
  {author} {\bibfnamefont {R.}~\bibnamefont {Ozeri}},\ }\bibfield  {title}
  {\bibinfo {title} {{Direct reconstruction of the quantum-master-equation
  dynamics of a trapped-ion qubit}},\ }\href
  {https://doi.org/10.1103/PhysRevA.101.062305} {\bibfield  {journal} {\bibinfo
   {journal} {Physical Review A}\ }\textbf {\bibinfo {volume} {101}},\ \bibinfo
  {pages} {062305} (\bibinfo {year} {2020})}\BibitemShut {NoStop}%
\bibitem [{\citenamefont {Weimer}\ \emph {et~al.}(2010)\citenamefont {Weimer},
  \citenamefont {M{\"{u}}ller}, \citenamefont {Lesanovsky}, \citenamefont
  {Zoller},\ and\ \citenamefont {B{\"{u}}chler}}]{Weimer2010}%
  \BibitemOpen
  \bibfield  {author} {\bibinfo {author} {\bibfnamefont {H.}~\bibnamefont
  {Weimer}}, \bibinfo {author} {\bibfnamefont {M.}~\bibnamefont
  {M{\"{u}}ller}}, \bibinfo {author} {\bibfnamefont {I.}~\bibnamefont
  {Lesanovsky}}, \bibinfo {author} {\bibfnamefont {P.}~\bibnamefont {Zoller}},\
  and\ \bibinfo {author} {\bibfnamefont {H.~P.}\ \bibnamefont
  {B{\"{u}}chler}},\ }\bibfield  {title} {\bibinfo {title} {{A Rydberg quantum
  simulator}},\ }\href {https://doi.org/10.1038/nphys1614} {\bibfield
  {journal} {\bibinfo  {journal} {Nature Physics}\ }\textbf {\bibinfo {volume}
  {6}},\ \bibinfo {pages} {382} (\bibinfo {year} {2010})}\BibitemShut {NoStop}%
\bibitem [{\citenamefont {Pop}\ \emph {et~al.}(2014)\citenamefont {Pop},
  \citenamefont {Geerlings}, \citenamefont {Catelani}, \citenamefont
  {Schoelkopf}, \citenamefont {Glazman},\ and\ \citenamefont
  {Devoret}}]{Pop2014}%
  \BibitemOpen
  \bibfield  {author} {\bibinfo {author} {\bibfnamefont {I.~M.}\ \bibnamefont
  {Pop}}, \bibinfo {author} {\bibfnamefont {K.}~\bibnamefont {Geerlings}},
  \bibinfo {author} {\bibfnamefont {G.}~\bibnamefont {Catelani}}, \bibinfo
  {author} {\bibfnamefont {R.~J.}\ \bibnamefont {Schoelkopf}}, \bibinfo
  {author} {\bibfnamefont {L.~I.}\ \bibnamefont {Glazman}},\ and\ \bibinfo
  {author} {\bibfnamefont {M.~H.}\ \bibnamefont {Devoret}},\ }\bibfield
  {title} {\bibinfo {title} {Coherent suppression of electromagnetic
  dissipation due to superconducting quasiparticles},\ }\href
  {https://doi.org/10.1038/nature13017} {\bibfield  {journal} {\bibinfo
  {journal} {Nature}\ }\textbf {\bibinfo {volume} {508}},\ \bibinfo {pages}
  {369} (\bibinfo {year} {2014})}\BibitemShut {NoStop}%
\bibitem [{\citenamefont {Earnest}\ \emph {et~al.}(2018)\citenamefont
  {Earnest}, \citenamefont {Chakram}, \citenamefont {Lu}, \citenamefont
  {Irons}, \citenamefont {Naik}, \citenamefont {Leung}, \citenamefont {Ocola},
  \citenamefont {Czaplewski}, \citenamefont {Baker}, \citenamefont {Lawrence},
  \citenamefont {Koch},\ and\ \citenamefont {Schuster}}]{Earnest2018}%
  \BibitemOpen
  \bibfield  {author} {\bibinfo {author} {\bibfnamefont {N.}~\bibnamefont
  {Earnest}}, \bibinfo {author} {\bibfnamefont {S.}~\bibnamefont {Chakram}},
  \bibinfo {author} {\bibfnamefont {Y.}~\bibnamefont {Lu}}, \bibinfo {author}
  {\bibfnamefont {N.}~\bibnamefont {Irons}}, \bibinfo {author} {\bibfnamefont
  {R.~K.}\ \bibnamefont {Naik}}, \bibinfo {author} {\bibfnamefont
  {N.}~\bibnamefont {Leung}}, \bibinfo {author} {\bibfnamefont
  {L.}~\bibnamefont {Ocola}}, \bibinfo {author} {\bibfnamefont {D.~A.}\
  \bibnamefont {Czaplewski}}, \bibinfo {author} {\bibfnamefont
  {B.}~\bibnamefont {Baker}}, \bibinfo {author} {\bibfnamefont
  {J.}~\bibnamefont {Lawrence}}, \bibinfo {author} {\bibfnamefont
  {J.}~\bibnamefont {Koch}},\ and\ \bibinfo {author} {\bibfnamefont {D.~I.}\
  \bibnamefont {Schuster}},\ }\bibfield  {title} {\bibinfo {title} {Realization
  of a {{$\Lambda$ System}} with {{Metastable States}} of a {{Capacitively
  Shunted Fluxonium}}},\ }\href
  {https://doi.org/10.1103/PhysRevLett.120.150504} {\bibfield  {journal}
  {\bibinfo  {journal} {Phys. Rev. Lett.}\ }\textbf {\bibinfo {volume} {120}},\
  \bibinfo {pages} {150504} (\bibinfo {year} {2018})}\BibitemShut {NoStop}%
\bibitem [{\citenamefont {Hazard}\ \emph {et~al.}(2019)\citenamefont {Hazard},
  \citenamefont {Gyenis}, \citenamefont {Di~Paolo}, \citenamefont {Asfaw},
  \citenamefont {Lyon}, \citenamefont {Blais},\ and\ \citenamefont
  {Houck}}]{Hazard2019}%
  \BibitemOpen
  \bibfield  {author} {\bibinfo {author} {\bibfnamefont {T.~M.}\ \bibnamefont
  {Hazard}}, \bibinfo {author} {\bibfnamefont {A.}~\bibnamefont {Gyenis}},
  \bibinfo {author} {\bibfnamefont {A.}~\bibnamefont {Di~Paolo}}, \bibinfo
  {author} {\bibfnamefont {A.~T.}\ \bibnamefont {Asfaw}}, \bibinfo {author}
  {\bibfnamefont {S.~A.}\ \bibnamefont {Lyon}}, \bibinfo {author}
  {\bibfnamefont {A.}~\bibnamefont {Blais}},\ and\ \bibinfo {author}
  {\bibfnamefont {A.~A.}\ \bibnamefont {Houck}},\ }\bibfield  {title} {\bibinfo
  {title} {Nanowire {{Superinductance Fluxonium Qubit}}},\ }\href
  {https://doi.org/10.1103/PhysRevLett.122.010504} {\bibfield  {journal}
  {\bibinfo  {journal} {Phys. Rev. Lett.}\ }\textbf {\bibinfo {volume} {122}},\
  \bibinfo {pages} {010504} (\bibinfo {year} {2019})}\BibitemShut {NoStop}%
\bibitem [{\citenamefont {Saffman}\ \emph {et~al.}(2010)\citenamefont
  {Saffman}, \citenamefont {Walker},\ and\ \citenamefont
  {M{\o}lmer}}]{Saffman2010}%
  \BibitemOpen
  \bibfield  {author} {\bibinfo {author} {\bibfnamefont {M.}~\bibnamefont
  {Saffman}}, \bibinfo {author} {\bibfnamefont {T.~G.}\ \bibnamefont
  {Walker}},\ and\ \bibinfo {author} {\bibfnamefont {K.}~\bibnamefont
  {M{\o}lmer}},\ }\bibfield  {title} {\bibinfo {title} {{Quantum information
  with Rydberg atoms}},\ }\href {https://doi.org/10.1103/RevModPhys.82.2313}
  {\bibfield  {journal} {\bibinfo  {journal} {Reviews of Modern Physics}\
  }\textbf {\bibinfo {volume} {82}},\ \bibinfo {pages} {2313} (\bibinfo {year}
  {2010})}\BibitemShut {NoStop}%
\bibitem [{\citenamefont {Madjarov}\ \emph {et~al.}(2020)\citenamefont
  {Madjarov}, \citenamefont {Covey}, \citenamefont {Shaw}, \citenamefont
  {Choi}, \citenamefont {Kale}, \citenamefont {Cooper}, \citenamefont
  {Pichler}, \citenamefont {Schkolnik}, \citenamefont {Williams},\ and\
  \citenamefont {Endres}}]{Madjarov2020}%
  \BibitemOpen
  \bibfield  {author} {\bibinfo {author} {\bibfnamefont {I.~S.}\ \bibnamefont
  {Madjarov}}, \bibinfo {author} {\bibfnamefont {J.~P.}\ \bibnamefont {Covey}},
  \bibinfo {author} {\bibfnamefont {A.~L.}\ \bibnamefont {Shaw}}, \bibinfo
  {author} {\bibfnamefont {J.}~\bibnamefont {Choi}}, \bibinfo {author}
  {\bibfnamefont {A.}~\bibnamefont {Kale}}, \bibinfo {author} {\bibfnamefont
  {A.}~\bibnamefont {Cooper}}, \bibinfo {author} {\bibfnamefont
  {H.}~\bibnamefont {Pichler}}, \bibinfo {author} {\bibfnamefont
  {V.}~\bibnamefont {Schkolnik}}, \bibinfo {author} {\bibfnamefont {J.~R.}\
  \bibnamefont {Williams}},\ and\ \bibinfo {author} {\bibfnamefont
  {M.}~\bibnamefont {Endres}},\ }\bibfield  {title} {\bibinfo {title}
  {{High-fidelity entanglement and detection of alkaline-earth Rydberg
  atoms}},\ }\href {https://doi.org/10.1038/s41567-020-0903-z} {\bibfield
  {journal} {\bibinfo  {journal} {Nature Physics}\ }\textbf {\bibinfo {volume}
  {16}},\ \bibinfo {pages} {857} (\bibinfo {year} {2020})}\BibitemShut
  {NoStop}%
\bibitem [{\citenamefont {Gan}\ \emph {et~al.}(2020)\citenamefont {Gan},
  \citenamefont {Maslennikov}, \citenamefont {Tseng}, \citenamefont {Nguyen},\
  and\ \citenamefont {Matsukevich}}]{Gan2020}%
  \BibitemOpen
  \bibfield  {author} {\bibinfo {author} {\bibfnamefont {H.~C.~J.}\
  \bibnamefont {Gan}}, \bibinfo {author} {\bibfnamefont {G.}~\bibnamefont
  {Maslennikov}}, \bibinfo {author} {\bibfnamefont {K.-W.}\ \bibnamefont
  {Tseng}}, \bibinfo {author} {\bibfnamefont {C.}~\bibnamefont {Nguyen}},\ and\
  \bibinfo {author} {\bibfnamefont {D.}~\bibnamefont {Matsukevich}},\
  }\bibfield  {title} {\bibinfo {title} {{Hybrid Quantum Computing with
  Conditional Beam Splitter Gate in Trapped Ion System}},\ }\href
  {https://doi.org/10.1103/PhysRevLett.124.170502} {\bibfield  {journal}
  {\bibinfo  {journal} {Physical Review Letters}\ }\textbf {\bibinfo {volume}
  {124}},\ \bibinfo {pages} {170502} (\bibinfo {year} {2020})}\BibitemShut
  {NoStop}%
\bibitem [{\citenamefont {Wang}\ \emph {et~al.}(2020)\citenamefont {Wang},
  \citenamefont {Lai}, \citenamefont {Yuan}, \citenamefont {Suh},\ and\
  \citenamefont {Vahala}}]{Wang2020}%
  \BibitemOpen
  \bibfield  {author} {\bibinfo {author} {\bibfnamefont {H.}~\bibnamefont
  {Wang}}, \bibinfo {author} {\bibfnamefont {Y.-H.}\ \bibnamefont {Lai}},
  \bibinfo {author} {\bibfnamefont {Z.}~\bibnamefont {Yuan}}, \bibinfo {author}
  {\bibfnamefont {M.-G.}\ \bibnamefont {Suh}},\ and\ \bibinfo {author}
  {\bibfnamefont {K.}~\bibnamefont {Vahala}},\ }\bibfield  {title} {\bibinfo
  {title} {{Petermann-factor sensitivity limit near an exceptional point in a
  Brillouin ring laser gyroscope}},\ }\href
  {https://doi.org/10.1038/s41467-020-15341-6} {\bibfield  {journal} {\bibinfo
  {journal} {Nature Communications}\ }\textbf {\bibinfo {volume} {11}},\
  \bibinfo {pages} {1610} (\bibinfo {year} {2020})}\BibitemShut {NoStop}%
\bibitem [{\citenamefont {Khandelwal}\ \emph {et~al.}(2021)\citenamefont
  {Khandelwal}, \citenamefont {Brunner},\ and\ \citenamefont
  {Haack}}]{Khandelwal2021b}%
  \BibitemOpen
  \bibfield  {author} {\bibinfo {author} {\bibfnamefont {S.}~\bibnamefont
  {Khandelwal}}, \bibinfo {author} {\bibfnamefont {N.}~\bibnamefont
  {Brunner}},\ and\ \bibinfo {author} {\bibfnamefont {G.}~\bibnamefont
  {Haack}},\ }\bibfield  {title} {\bibinfo {title} {{Signatures of exceptional
  points in a quantum thermal machine}},\ }\href
  {http://arxiv.org/abs/2101.11553} {\ ,\ \bibinfo {pages} {1} (\bibinfo {year}
  {2021})},\ \Eprint {https://arxiv.org/abs/2101.11553} {arXiv:2101.11553}
  \BibitemShut {NoStop}%
\end{thebibliography}%

\clearpage
\renewcommand{\thefigure}{S\arabic{figure}}
\renewcommand{\figurename}{Supplemental Figure}
\setcounter{figure}{0}
\onecolumngrid
\section*{SUPPLEMENTARY INFORMATION}
The main results of the manuscript ``Optimized Quantum Steering and
Exceptional Points'' discuss the state steering of a two-level system
towards an arbitrary desired target state and the optimization of
the convergence rate. In the manuscript, the convergence rate optimization
is linked to the exceptional points. Here we provide supplementary
material for the results in the manuscript. In Section \ref{sec:Lower-bound-on-Omega},
we obtain the lower bound on allowed $\Omega$ for a fixed target
state. In Section \ref{sec:Optimal-convergence-rate-and-Omega-medium-purity},
we present details on the optimal convergence rates in the three purity
regimes.

\section{Lower bound on $\Omega$~\label{sec:Lower-bound-on-Omega}}

In the manuscript, we discuss that there exist a range of $\Omega$
to choose from for a given target state, and we vary $\Omega$ to
maximize the convergence rate. Here we will show that the allowed
values of $\Omega$ required for a given target state has a lower
bound. For a given detector state initialization direction $\hat{m}=(0,0,1)$,
the Bloch coordinates of the system's steady state are given by

\begin{eqnarray}
s_{x} & = & \frac{2\Omega\sin\theta(\Omega\cos\theta\cos\phi+\sin\phi)}{2+\Omega^{2}(\cos^{2}\theta+1)},\label{eq:sx_steady_state_coordinate}\\
s_{y} & = & \frac{2\Omega\sin\theta(\Omega\cos\theta\sin\phi-\cos\phi)}{2+\Omega^{2}(\cos^{2}\theta+1)},\label{eq:sy_steady_state_coordinate}\\
s_{z} & = & \frac{2(1+\Omega^{2}\cos^{2}\theta)}{2+\Omega^{2}(\cos^{2}\theta+1)},\label{eq:sz_steady_state_coordinate}
\end{eqnarray}
where $\Omega=\omega/\alpha$. We note that the steady state given
by Eqs.~(\ref{eq:sx_steady_state_coordinate},\ref{eq:sy_steady_state_coordinate},\ref{eq:sz_steady_state_coordinate})
lies on an ellipsoid as

\begin{equation}
2(s_{x}^{2}+s_{y}^{2})+4\left(s_{z}-\frac{1}{2}\right)^{2}=1.\label{eq:ellipsoidal equation}
\end{equation}
Using Eqs.~(\ref{eq:sx_steady_state_coordinate}) and (\ref{eq:sy_steady_state_coordinate}),
we get

\begin{equation}
\Omega=\frac{(s_{x}+s_{y}\tan\phi)}{\cos\theta\,(s_{x}\tan\phi-s_{y})}.\label{eq:Omega_from_sx_and_sy}
\end{equation}
Then from Eqs.~(\ref{eq:sz_steady_state_coordinate}) and (\ref{eq:Omega_from_sx_and_sy}),
we get

\begin{equation}
\Omega^{2}=\frac{(s_{x}+s_{y}\tan\phi)^{2}-(s_{x}^{2}+s_{y}^{2})+2(1-s_{z})(s_{z}+2(1-s_{z})\sec^{2}\phi)}{(s_{x}\tan\phi-s_{y})^{2}}.\label{eq:Omega_free_from_theta}
\end{equation}
Therefore, for a given steady state, $\Omega$ becomes a function
of $\phi$ only. Assuming $\Omega\ge0$, we see that there exist only
one extremum with respect to $\phi$ (which is a minimum), and the minimum
value of $\Omega$ is given by

\begin{equation}
\Omega_{\textrm{min}}=\sqrt{\frac{2(1-s_{z})}{s_{z}}}\quad\quad\quad\quad\textrm{at}\quad\quad\quad\quad\phi=\phi_{c}=\tan^{-1}\left(-\frac{s_{x}}{s_{y}}\right).\label{eq:Omega_min_and_phi_critical}
\end{equation}
For a given steady state on the ellipsoid, the Bloch coordinate $s_{z}$
is related to the state purity $\purity$ as

\begin{equation}
s_{z}=1-\sqrt{2(1-\purity)}.\label{eq:relation_between_sz_and_purity}
\end{equation}
Therefore, from Eq.~(\ref{eq:Omega_min_and_phi_critical}), we get

\begin{equation}
\Omega_{\textrm{min}}=\sqrt{\frac{2\sqrt{2(1-\purity)}}{1-\sqrt{2(1-\purity)}}}.\label{eq:Omega_min_in_terms_of_purity}
\end{equation}
This implies that there exist a lower bound on the allowed values
of $\Omega$ required to steer a system towards a desired target state
and $\Omega_{\textrm{min}}$ increases as target state purity decreases.
Since $\Omega=\omega/\alpha$, this means one needs stronger Zeeman
field in order to generate a lower purity target state. 

\section{Optimal convergence rates ~\label{sec:Optimal-convergence-rate-and-Omega-medium-purity}}

In this section, we present details on the optimal convergence rates
in the three purity regimes.

\subsection{Eigenvalues of the Liouvillian superoperator}

In this subsection, we discuss details on how to compute the eigenvalues
of the Liouvillian superator. In the main manuscript, Eq.~(\ref{eq:master equation}) defines
the Liouvillian superoperator $\mathcal{L}$ of our system.

The Liouvillian superoperators for two different detector state initialization
directions, say $\hat{m}$ and $\hat{n}$, are related by a unitary
transform. Specifically, let $U$ denote the unitary transform that
rotates an eigenstate for one direction, $|\hat{m}_{+}\rangle$, to
an eigenstate for the other direction, $|\hat{n}_{+}\rangle=U|\hat{m}_{+}\rangle$.
Then, the Liouvillian superoperators for the two directions, denoted
$\mathcal{L}_{\hat{m}}$ and $\mathcal{L}_{\hat{n}}$, are related
by $\mathcal{L}_{\hat{n}}[\rho]=U\mathcal{L}_{\hat{m}}[U^{\dagger}\rho U]U^{\dagger}$.
This fact implies that the superoperators have the same eigenvalues,
because if $\rho^{(j)}$ is an eigenvector of $\mathcal{L}_{\hat{m}}$,
then $\tilde{\rho}^{(j)}=U\rho^{(j)}U^{\dagger}$ is an eigenvector
of $\mathcal{L}_{\hat{n}}$ with the same eigenvalue. Therefore, for
simplicity, we choose $\hat{m}=(0,0,1)$.

For the detector state initialization direction $\hat{m}=(0,0,1)$,
the Liouvillian can be expressed as a $4\times4$-matrix acting on
the space of density matrices:

\begin{equation}
\mathcal{L}=\alpha\left(\begin{array}{cccc}
0 & i\,e^{i\phi}\Omega\,\sin\theta & -i\,e^{-i\phi}\Omega\,\sin\theta & 4\\
i\,e^{-i\phi}\Omega\,\sin\theta & -2(i\Omega\,\cos\theta+1) & 0 & -i\,e^{-i\phi}\Omega\,\sin\theta\\
-i\,e^{i\phi}\Omega\,\sin\theta & 0 & 2(i\Omega\,\cos\theta-1) & i\,e^{i\phi}\Omega\,\sin\theta\\
0 & -i\,e^{i\phi}\Omega\,\sin\theta & i\,e^{-i\phi}\Omega\,\sin\theta & -4
\end{array}\right).\label{eq:Liouvillian superoperator}
\end{equation}
The eigenvalues of the superoperator are the zeros of the characteristic
polynomial $\det(\lambda\mathbb{I}-\mathcal{L}$) of $\mathcal{L}$.
Using that the steady state corresponds to the eigenvalue $\lambda=0$,
we can simplify the characteristic polynomial and find that the nonzero
eigenvalues satisfy the polynomial equation

\begin{equation}
\mathcal{C}(\lambda)=\lambda^{3}+8\alpha\lambda^{2}+4\alpha^{2}\left(5+\Omega^{2}\right)\lambda+8\alpha^{3}\left(2+\Omega^{2}(1+\cos^{2}\theta)\right)=0.\label{eq:char_poly_1}
\end{equation}
Substituting $\cos^{2}\theta$ from Eq.~(\ref{eq:sz_steady_state_coordinate}),
and rescaling $\lambda$ as $\lambda=\alpha\,\Lambda$, we can write
Eq.~(\ref{eq:char_poly_1}) as

\begin{equation}
\mathcal{C}(\Lambda)=\Lambda^{3}+8\Lambda^{2}+4(5+\Omega^{2})\Lambda+16\left(\frac{1+\Omega^{2}}{1+\sqrt{2(1-\purity)}}\right)=0,\label{eq:char_poly_2}
\end{equation}
where we have used $s_{z}=1-\sqrt{2(1-\purity)}$.

We note that the eigenvalues of $\mathcal{L}$ consist of isolated
real numbers and pairs of complex conjugate numbers. This fact holds
because the coefficients of the characteristic polynomial are real
quantities; thus, if $\lambda$ is an eigenvalue of the Liouvillian
superoperator, then its complex conjugate $\lambda^{*}$ is also an
eigenvalue. More generally, this fact is a consequence of the fact
that the time evolution preserves the Hermiticity of the density matrix,
which implies that $\mathcal{L}[\rho_{s}^{\dagger}]=\left(\mathcal{L}[\rho_{s}]\right)^{\dagger}$.

\subsection{Low purity regime}

In the low purity regime, we find numerically that one nonzero eigenvalue
of the Liouvillian $\mathcal{L}$ is real, while the other two form
a complex conjugate pair. The magnitude of the real eigenvalue increases
monotonically as $\Omega$ increases. The real eigenvalue always corresponds
to the slowest decay rate.

\subsection{Medium purity regime}

In this subsection, we discuss the medium purity regime and show that
there is a critical value $\Omega$ where all three nonzero eigenvalues
of the Liouvillian superoperator have equal real parts, and that the
convergence rate becomes optimal at this value.

We find numerically that one nonzero eigenvalue of the Liouvillian
$\mathcal{L}$ is real, while the other two eigenvalues form a complex
conjugate pair. Furthermore, the magnitude of the real eigenvalues
increases monotonically, whereas the magnitude of the real parts of
the complex eigenvalues decreases monotonically. In contrast to the
low purity regime, the three real parts of the eigenvalues do become
equal for some critical value of $\Omega$ in the medium purity regime.
To find this critical value analytically, we make the ansatz

\begin{equation}
\Lambda_{1}=a,\,\Lambda_{2}=a+i\,b,\,\Lambda_{3}=a-i\,b
\end{equation}
for the eigenvalues, where $a$ denotes the common real part, and
$b$ denotes the imaginary part of the complex conjugate pair. Then,
the characteristic polynomial can be factored as

\begin{equation}
\mathcal{C}(\Lambda)=(\Lambda-a)(\Lambda-a-i\,b)(\Lambda-a+i\,b).\label{eq:char_poly_med_regime}
\end{equation}
Expanding this expression and comparing it with Eq.~(\ref{eq:char_poly_2}),
we can solve for $a,\,b$ and $\Omega$, obtaining

\begin{equation}
a=-\frac{8}{3},\,b=\pm\frac{4}{3}\sqrt{\frac{8\sqrt{2(1-\purity)}-1}{1-2\sqrt{2(1-\purity)}}},\ \Omega=\pm\sqrt{\frac{26\sqrt{2(1-\purity)}-1}{9(1-2\sqrt{2(1-\purity)})}}\,.\label{eq:med_purity_a_b_Omega}
\end{equation}
Since the convergence rate is defined as the magnitude of the real
part of the eigenvalues, we obtain $\Gamma=8\alpha/3$ for the convergence
rate at this value of $\Omega$. We have argued in the main text that
the equality of the real parts implies that this convergence rate
is optimal.

The range of the medium purity regime is determined by the constraint
that the quantities $b$ and $\Omega$ have to be real. We find that
this is satisfied precisely for $7/8<\purity\le127/128$. At $\purity=127/128$,
we have $b=0$, i.~e.~all three nonzero eigenvalues becomes purely
real and equal to each other, implying that the Liouvillian superoperator
features a third-order exceptional point which marks the onset of
the high purity regime.

\subsection{High purity regime}

In the high purity regime, we find numerically that the optimal convergence
rate occurs at a critical value of $\Omega$ where the Liouvillian
superoperator has a second-order exceptional point, that is where
two of its eigenvalues coincide. To compute the exceptional points
analytically, we follow a procedure similar to the procedure in the
previous subsection. Specifically, we make the ansatz

\begin{equation}
\Lambda_{1}=\Lambda_{2}=a,\,\Lambda_{3}=b
\end{equation}
for the eigenvalues of the Liouvillian $\mathcal{L}$. Since the eigenvalues
are either real or have complex conjugate pairs, the equality of two
of them implies that $a$ and $b$ are real quantities. The characteristic
equation can be factored as

\begin{equation}
\mathcal{C}(\Lambda)=(\Lambda-a)^{2}(\Lambda-b)=0.\label{eq:char_poly_high_purity_regime}
\end{equation}
Expanding this expression and comparing with Eq.~(\ref{eq:char_poly_2}),
we obtain two exceptional points:

\begin{equation}
a=a_{+},\,b=b_{-}\,\Omega=\Omega_{+}\label{eq:soln_set_1_in_high_purity_regime}
\end{equation}
and 
\begin{equation}
a=a_{-},\,b=b_{+},\,\Omega=\Omega_{-},\label{eq:soln_set_2_in_high_purity_regime}
\end{equation}
where

\begin{eqnarray}
a_{\pm} & = & \frac{-3\pm\sqrt{1-8\sqrt{2(1-\purity})}}{1+\sqrt{2(1-\purity)}},\label{eq:a_in_high_purity_regime}\\
b_{\pm} & = & \frac{2\left(-1-4\sqrt{2(1-\purity)}\pm\sqrt{1-8\sqrt{2(1-\purity)}}\right)}{1+\sqrt{2(1-\purity)}},\label{eq:b_in_high_purity_regime}\\
\Omega_{\pm} & = & \sqrt{\frac{16\sqrt{2(1-\purity)}+20\purity-21\pm\left(1-8\sqrt{2(1-\purity)}\right)^{3/2}}{2\left(1+\sqrt{2(1-\purity)}\right)^{2}}}.\label{eq:Omega_in_high_purity_regime}
\end{eqnarray}
However, we note that for $161/162<\purity<1$, the exceptional point
at $\Omega_{-}$ falls outside the admissible range for $\Omega$
computed in Section~\ref{sec:Lower-bound-on-Omega}, that is $\Omega_{-}<\Omega_{\textrm{min}}$.
Numerically, we find that the convergence rate becomes optimal at
$\Omega=\Omega_{+}$, and is given by $|a_{+}|$.
\end{document}